\documentclass[a4paper,11pt]{article}
\pdfoutput=1
\usepackage{jheppub}
\usepackage[utf8]{inputenc}

\usepackage{cleveref}
\crefformat{figure}{figure~#2#1#3}
\crefformat{table}{table~#2#1#3}
\crefformat{equation}{eq.~(#2#1#3)}
\crefrangeformat{equation}{eqs.~(#3#1#4) to~(#5#2#6)}
\crefmultiformat{equation}{eqs.~(#2#1#3)}{ and~(#2#1#3)}{, (#2#1#3)}{ and~(#2#1#3)}
\crefrangemultiformat{equation}{eqs.~(#2#1#3)}{ and~(#2#1#3)}{, (#2#1#3)}{ and~(#2#1#3)}

\newcommand{\be}{\begin{equation}}
\newcommand{\ee}{\end{equation}}
\newcommand{\nn}{\nonumber}
\newcommand{\Gam}{\Gamma}
\newcommand{\eps}{\epsilon}
\newcommand{\pbar}{\bar{p}}
\global\long\def\order#1{\mathcal{O}\left(#1\right)}

\title{Double-real contribution to the quark beam function at N$^3$LO QCD}

\author[a]{K. Melnikov,}
\author[a]{R. Rietkerk,}
\author[b]{L. Tancredi,}
\author[a,c]{and C. Wever}

\affiliation[a]{Institute for Theoretical Particle Physics, KIT, Karlsruhe, Germany}
\affiliation[b]{Theoretical Physics Department, CERN, 1211 Geneva 23, Switzerland}
\affiliation[c]{Institut f\"ur Kernphysik, KIT, 76344 Eggenstein-Leopoldshafen, Germany}

\emailAdd{kirill.melnikov@kit.edu}
\emailAdd{robbert.rietkerk@kit.edu}
\emailAdd{lorenzo.tancredi@cern.ch}
\emailAdd{christopher.wever@kit.edu}

\abstract{
We compute the master integrals required for the calculation of the double-real emission contributions to the matching coefficients of 0-jettiness beam functions at next-to-next-to-next-to-leading order in perturbative QCD.  
As an application, we combine these integrals and derive the double-real gluon emission contribution to the matching coefficient $I_{qq}(t,z)$ of the quark beam function. 
}


\arxivnumber{1809.06300}

\begin{document} 
\maketitle
\flushbottom

\allowdisplaybreaks

\section{Introduction } 
\label{sec:intro}

The absence of any evidence for   physics beyond the Standard Model 
at the LHC implies a growing importance  of indirect searches 
for new particles and interactions. An integral part of this complex 
endeavour are first-principles predictions for hard scattering processes
in proton collisions  with controllable perturbative accuracy.
In recent years, we have seen a remarkable progress in an effort 
to provide such predictions. 

Indeed, robust methods for one-loop  computations developed 
during the past decade, that allowed the theoretical description of a large 
number of processes with 
multi-particle final  states through NLO QCD \cite{Bern:1996je,Bern:1997sc,
Ossola:2006us,Berger:2008sj,Giele:2008ve,Badger:2008cm}, were followed by 
the development of practical NNLO QCD subtraction and slicing 
schemes \cite{GehrmannDeRidder:2005cm,Boughezal:2011jf,Catani:2007vq,Czakon:2010td,Czakon:2014oma,
Cacciari:2015jma,DelDuca:2015zqa,DelDuca:2016ily,Gaunt:2015pea,Boughezal:2015aha,Caola:2017xuq} 
and advances in computations of two-loop scattering amplitudes \cite{Chetyrkin:1981qh,Kotikov:1990kg,Bern:1993kr,Remiddi:1997ny,Remiddi:1999ew,Gehrmann:1999as,Laporta:2001dd,Vollinga:2004sn,Goncharov:2010jf,Duhr:2011zq,Duhr:2012fh,Henn:2013pwa}.
This progress led  to an  opportunity to describe many  $2 \to 2$ 
partonic processes relevant for the LHC physics 
with the NNLO QCD accuracy. 

These impressive  developments were recently extended  by  
a breakthrough  computation of the 
N$^3$LO QCD corrections to Higgs boson production in gluon fusion \cite{Anastasiou:2013srw, 
Dulat:2017prg,Mistlberger:2018etf}. 
Both, the total cross section and simple kinematic distributions were computed
in these references.  
The computational  techniques employed there 
rely  heavily on the use of reverse unitarity 
\cite{Anastasiou:2002yz}
that  allows one to map complex phase space integrals 
to loop integrals and use the machinery 
of multi-loop computations to reduce the number of independent integrals that need to be computed.

It is clear that further  applications of this technology will enable  
the computation of 
$Z$ and $W$ production cross sections and basic kinematic distributions 
through N$^3$LO QCD as well.  
However,  it should be also recognized that the 
theoretical methods employed in refs.~\cite{Anastasiou:2013srw,Dulat:2017prg} limit the number of observables 
that can  be studied with  such 
high-order perturbative accuracy. Indeed, it is highly unlikely 
that complex fiducial cross  sections defined at the level of decay products 
of the produced color-singlet particles, 
with additional  restrictions on the QCD radiation, can be computed using these techniques. 
Yet, it is the knowledge of these fiducial cross sections that 
allows a direct connection 
of the refined theoretical predictions with  the results of the experimental measurements. 
For this reason, the extension of the results of  
refs.~\cite{Anastasiou:2013srw,Dulat:2017prg} towards fully-differential cross sections and 
distributions is highly desirable. 

It is known from NNLO QCD computations that the calculation  
of an arbitrary fiducial cross section  or kinematic 
distribution  requires the development of a full-fledged  subtraction scheme 
to identify and remove   infra-red and collinear divergences. Developing 
such a subtraction scheme 
at N$^3$LO QCD is a daunting task.  An alternative, more practical,  approach 
is to use  slicing methods known from NLO QCD \cite{Giele:1991vf}
 and extended recently to NNLO 
computations \cite{Catani:2007vq,Gaunt:2015pea,Boughezal:2015aha}.

To explain the general idea of the slicing method, we consider a process where a color-singlet final 
state $V$ is produced in proton-proton collisions together with some accompanying QCD 
radiation $X$. We do not require the presence of any jets in the final state. 
It is possible to choose an infra-red safe 
kinematic variable, that we will refer to as $\omega$, with the property 
that $\omega[V] = 0$ and $\omega[V,X] > 0$, provided  that the momenta 
of gluons and quarks in the final state $X$ are neither soft nor 
collinear to the incoming protons. 
It follows that,  if we consider the process $pp \to V+X$ and require that $\omega[V,X] > \omega_0 > 0$, 
we prevent   the final state QCD  partons from becoming  fully unresolved. From the 
viewpoint of  fixed order 
 perturbative computations, this implies that, for $\omega > \omega_0$, 
we consider a process $pp \to V+j$, so that  
the final color-singlet state $V$ is outside of the  Born phase space.  As the result, 
the desired N$^3$LO QCD computation becomes a NNLO QCD computation for $pp \to V+j$ 
for $\omega > \omega_0$.   Given the recent progress, such 
NNLO QCD computations are definitely possible. 

To enable the description of the inclusive process $pp \to V$ through N$^3$LO QCD, we need to supplement 
the NNLO QCD computation for $pp \to V+j$ described above with the computation of the contribution
to the inclusive cross section  from phase space region where 
$0 \le  \omega[V,X] < \omega_0$.  Note that this contribution includes 
$\omega[V,X]= 0$ and, therefore, 
is sensitive to the fully  unresolved kinematics. 
For finite $\omega_0$, the computation of the $\omega < \omega_0$ contribution 
is as difficult as the full computation. However, significant simplifications 
occur  if we take $\omega_0$ to be very small, $\omega_0 \ll 1$. For such 
small values of $\omega < \omega_0$, the allowed radiation $X$ is either 
soft or collinear. This already leads to certain simplifications of the required 
computations  
since, in soft and collinear limits, 
the matrix elements for the partonic processes 
factorize into 
hard process-dependent matrix elements and 
universal splitting  functions or  eikonal factors.
However, if this is  the end of the 
story, the required computations are still highly non-trivial 
because soft and collinear divergences do overlap. 
Luckily, this problem
can be ameliorated by  choosing a particular slicing  variable $\omega$.
Indeed, for certain choices of 
the slicing variables, there are factorization theorems that express 
cross sections at small values of $\omega$ 
through simpler objects  that separate 
soft and collinear dynamics and simplify 
the relevant computations significantly. 

This is exactly what happens for the so-called 0-jettiness variable 
\cite{Stewart:2009yx,Stewart:2010tn}, whose general factorization formula was originally derived  in SCET~\cite{Bauer:2000ew,Bauer:2000yr,Bauer:2001ct,Bauer:2001yt,Bauer:2002nz}.
For the  inclusive production of a color-singlet final state in hadron 
collisions, without requiring any 
additional jets in the final state, the relevant 0-jettiness  
variable reads 
\be
\label{eq1}
{\cal T} = \sum_{j} {\rm min}_{i \in \{1,2\}} \left [ \frac{2 p_i \cdot k_j}{Q_i} \right ].
\ee
Here,  $p_{1,2}$ are the momenta of the incoming protons, taken to be light-like, 
$Q_{1,2}$ are hardness variables 
that can be chosen in a number of different ways,   
and $k_{1,2,...N}$ are the momenta of the QCD partons that appear in the final state 
of the process $pp \to V + X$.

The usefulness of ${\cal T}$ as the slicing variable follows from the factorization theorem \cite{Stewart:2009yx} that states that for small values of ${\cal T}$ the differential cross section for $pp \to V+X$ can be written as the convolution of the so-called beam, soft and hard functions and the born cross section for $pp \to V$.
Schematically,
\be
\label{eq2}
\lim_{{\cal T} \to 0} {\rm d} \sigma(pp \to V+X) = B \otimes B \otimes S \otimes H \otimes {\rm d} \sigma(pp \to V)\,.
\ee
The soft function can be computed order-by-order in perturbation theory. On the contrary, 
the beam function is determined through a convolution of perturbatively calculable 
matching coefficient and the non-perturbative parton distribution functions \cite{Stewart:2009yx,Stewart:2010qs}
\be
B_{i}(t,x,\mu)  = \sum_{j \in {q,\bar q,g}} \int {\rm d} x_1 {\rm d} x_2 \, I_{ij}(t,x_1,\mu) f_j(x_2,\mu^2) \delta(x - x_1 x_2)\,. 
\label{eq:defBeamF}
\ee
The quantity $t$ in this formula is the so-called transverse  virtuality of the quark that  
participates in a hard process; it is related to the 0-jettiness variable by a simple 
rescaling. 
The index $i$ runs over all possible partons including quarks, anti-quarks and gluons. 

To arrive at the final result for the cross section in \cref{eq2} at a particular order in perturbation theory, 
 we need to compute the  matching coefficients 
$I_{ij}(t,x_1,\mu)$, $i,j \in \{q,\bar q,g\}$,
and the soft function  to  the same order in the perturbative expansion, subtract the
divergences by performing PDF renormalization and compute the relevant convolutions. 
Since we are interested in a fully-exclusive N$^3$LO computations, we need to know the 
soft and the beam functions  through N$^3$LO QCD  in perturbation theory. 

We will focus on the computation of the beam function at N$^3$LO QCD. 
At NLO, quark and gluon beam functions have been calculated in~\cite{Stewart:2010qs,Berger:2010xi},
while results at NNLO have been derived in~\cite{Gaunt:2014xga,Gaunt:2014cfa}.
The easiest way to think about the different ingredients 
of the computation is to realize that the matching coefficient $I_{ij}$ 
can be computed as a phase space integral
of  the collinear limit of the relevant scattering amplitudes squared. 
Integrations over  multi-particle  
phase spaces  are performed with 
constraints  on a transverse virtuality  $t$ 
and the light-cone component of the four-momentum 
of a parton that participates in  the  hard scattering 
process \cite{Ritzmann:2014mka}. 

We work in third-order perturbative QCD and focus on the quark-to-quark branching process with additional gluon emissions for definiteness.
We then need to consider various transitions of the type $q \to q^* + \{g_i\}$, 
where the number of additional gluons emitted to the final state changes from one to three. 
Additional powers of the strong coupling constant, required 
for a particular final state to contribute at  N$^3$LO,   are then provided by virtual  
diagrams that renormalize the collinear emissions of one and two real gluons. 

We now define the phase space constraint more precisely. To this end, we denote the four-momentum 
of the incoming quark as $p$, the four-momentum of a virtual quark $q^*$ that goes into the 
hard scattering as $p_*$ and the total momentum of the emitted gluons as $k_{\rm tot} = \sum \limits_{i=1}^{N} k_i$.
We fix the component of $p_*$ along the direction of the incoming momentum 
$p$ to be $z$. We then write
\be \label{eq4}
p_*^\mu = z p^\mu + y {\bar p}^\mu + k_\perp^\mu,
\;\;\; k_{\rm tot} = (1-z) p^\mu - y {\bar p}^\mu - k_\perp^\mu\,,
\ee
where ${\bar p}^2 = 0$ and $p \cdot k_\perp = \bar p \cdot k_\perp = 0$. 
We define the transverse  virtuality $t$ as
\be
t = -(p_*^2 - k_\perp^2) = -z y\,  2 p \cdot \bar p\,.
\ee
The value of 
$y$ can be computed by taking a  scalar product of $k_{\rm tot}$ with $p$, 
$y = - (p \cdot k_{\rm tot})/(p \cdot \bar p)$. 
We obtain 
\be
\label{eq6}
t = z \, 2 p \cdot k_{\rm tot} = z  \sum \limits_{i=1}^{N} 2 \, p \cdot k_{i}\,.
\ee

It is also useful to express the constraint on the light-cone component of the virtual 
quark momentum $p_*^{\mu}$ through 
the momenta of the emitted gluons. We do this by considering the scalar product of $p_*$ with $ \bar p$ 
and using momentum conservation $p^* = p - k_{\rm tot}$. Defining $s=2p \cdot \bar p$ we find 
\be
\label{eq7}
s (1-z) = \sum \limits_{i=1}^{N} 2 \bar p \cdot  k_{i}\,.
\ee

\Cref{eq6,eq7} provide the phase space constraints that need to be accounted 
for in the integration over the gluon phase space. Thus, the computation of the contribution 
of the $N$-gluon final 
state to the beam function is proportional to
\be
\label{eq8}
\begin{split}
I_N(t,z) &\sim  \int \prod \limits_{i=1}^{N}  \frac{{\rm d}^{d-1} k_i}{(2\pi)^{(d-1)} 2 k_i^0} 
\delta \left ( s(1-z) - \sum \limits_{i=1}^{N} 2 \bar p \cdot k_{i}  \right )  
\delta \left (t -   z \sum \limits_{i=1}^{N} 2 p \cdot k_{i}  \right ) \\
& \times \frac{ \hat C_p\left [ \left| {\cal M}(p,\bar p;\{k_i\}) \right|^2 \right ]}{ \left| {\cal M}_0(z p,\bar p) \right|^2} \,,
\end{split}
\ee
where  $\hat C_p$ denotes the collinear projection of the square of the full matrix element $|{\cal M}|^2$, 
following the recipe in ref.~\cite{Catani:1999ss}. 
The result is normalized to the square of tree-level matrix element $|{\cal M}_0|^2$. 
When working in a physical gauge all quantum effects reside on a single incoming quark line and emissions from the incoming anti-quark with momentum $\bar p$ decouple.

We will now discuss the three different contributions to the matching coefficient 
of the beam function through order ${\cal O}(\alpha_s^3)$.  The first one is the two-loop 
renormalization of a single collinear gluon emission. In this 
contribution the kinematics of a single real emission is fully 
constrained and the corresponding 
two-loop virtual correction  includes at most the two-loop 
three-point function with two 
light-like legs and one off-shell leg.  In principle, such contributions are known analytically if not
for the fact that they need to be computed  in the light-cone gauge to achieve collinear factorization.
The light-cone gauge introduces additional propagators to Feynman integrals making them unconventional
and difficult to compute. 

The second contribution is the single-virtual double-real emission process, i.e. a one-loop 
correction to the emission of two real gluons in the collinear kinematics. In this case 
the situation is more complex. Indeed, 
the one-loop virtual corrections include box diagrams 
that are sufficiently complicated and 
the kinematics of two real gluons is  sufficiently 
unconstrained to make the computation of this double-real contribution  
quite a challenging  task. The earlier comment about light-cone gauge also applies here. 

The last contribution is the triple real emission process, which involves the integration over
the  three-gluon 
phase space subject to the 0-jettiness constraint. Such a computation 
is also quite demanding.  
Finally, to arrive at the 
matching coefficient one has to perform collinear renormalization 
of the beam function which, at N$^3$LO, is also non-trivial.  

As the  result, we decided to report on this computation 
in a few separate  instalments. We will 
start with the discussion 
of  the single-virtual double-real contributions  to the matching 
coefficient of the beam function. It is 
schematically described by \cref{eq8} with  $N=2$.


\section{Matching coefficient}
\label{sec:amp}
In this section, we describe the calculation of the double-real contribution 
to the matching coefficients 
for the quark beam function at N$^3$LO, see \cref{eq:defBeamF}. 
We follow the observation made in ref.~\cite{Ritzmann:2014mka} 
and compute the matching coefficient as phase space 
integrals of  the corresponding splitting functions, using the appropriate 
kinematic constraints, see \cref{eq8}. 
The splitting functions at the relevant perturbative order 
can be constructed following the general method 
described in ref.~\cite{Catani:1999ss}.
The large number of phase-space integrals that appear in the course of the computation 
are calculated  using the method of reverse unitarity~\cite{Anastasiou:2002yz}.

\begin{figure}[t]
\begin{center}
\includegraphics[scale=0.55]{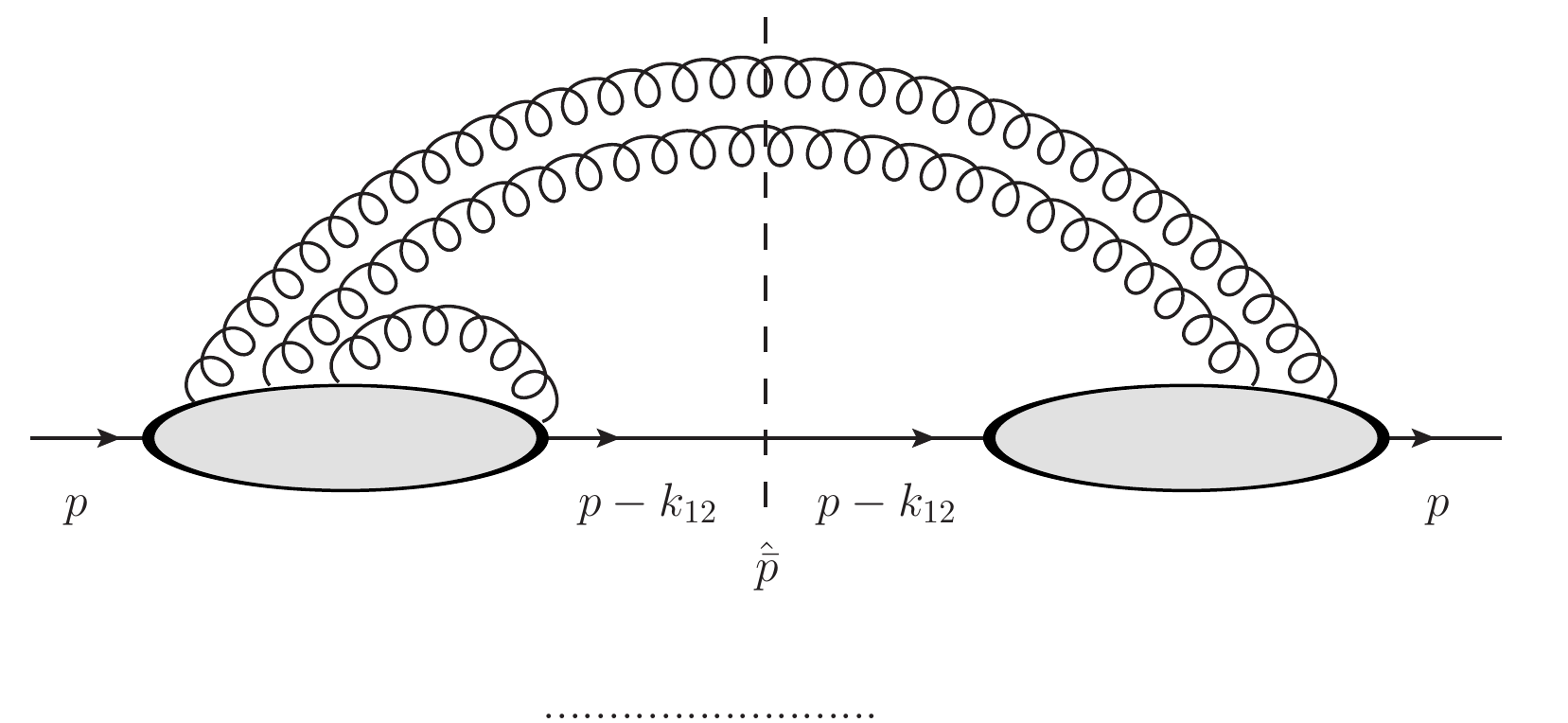}
\end{center}
\vspace{-1.1cm}
\caption{Double-real diagrams for the calculation of the quark beam function at N$^3$LO.}
\label{fig:1}
\end{figure}

We  consider a massless quark with  momentum $p$ that emits two
(collinear) gluons of momenta $k_1$ and $k_2$,
before entering the hard scattering process
$$q(p) \to q(q) + g(k_1) + g(k_2)\,, \qquad \mbox{with} \qquad q = p-k_1-k_2\,,$$
as depicted schematically in \cref{fig:1}.
We focus on  the one-loop
virtual corrections to the double gluon emission, i.e. 
a particular contribution to N$^3$LO beam function as we explained in the Introduction. 
Collinear dynamics on the quark line factorizes, 
which means that when computing the relevant 
amplitude squared in the collinear limit, 
we become insensitive to the hard scattering process. 
In ref.~\cite{Catani:1999ss}, it was shown that one can perform this projection 
by using a physical gauge for the real 
collinear gluons $g(k_1)$ and $g(k_2)$   and by inserting 
$\hat{\bar{p}} = \bar{p}^\mu \gamma_\mu$ in place of the 
hard scattering matrix element. The vector 
$\bar p$ is a light-cone vector complementary to the light-cone vector $p$, 
see \cref{eq4}. 

We find it practical to construct the collinear limit of the relevant amplitude 
squared in \cref{fig:1} by 
considering the Feynman diagrams for the process 
$q(p) \to q(q) + g(k_1) + g(k_2)\,.$
We employ  QGRAF~\cite{Nogueira:1991ex} to generate all tree-level and one-loop
diagrams, and use them to build up the amplitude in two independent implementations in
FORM~\cite{Vermaseren:2000nd} and Mathematica.
This requires  care, since the quark $q(p)$ is on-shell, while the quark $q(q)$ is off-shell. This means that when 
one-loop corrections are generated, 
we must  neglect all self-energy insertions on the on-shell external quark $q(p)$, but 
we need to include them on the off-shell quark $q(q)$.

As mentioned  above, we need to use light-cone gauge for both real and virtual gluons.
For example, when summing over real gluon polarizations, we obtain 
\begin{align}
\label{eq:pol}
\begin{split}
&\sum_{pol}\, \epsilon_i^\mu(k_i) \left( \epsilon_i^\nu( k_i) \right)^*  = 
  - g^{\mu \nu} + \frac{k_i^\mu \bar{p}^{\nu} + k_i^\nu \bar{p}^\mu}{k_i \cdot \bar{p}}\,,
  \;\;\; i = 1, 2\,.
\end{split}
\end{align}
Finally, we also need to use the axial gauge for virtual gluons, with the same reference vector $\pbar$.
We note that the choice of $\bar{p}$ as the reference vector is convenient but not necessary.

We find 3 tree-level diagrams and 30 one-loop diagrams. 
Computing the interference, we obtain 90 ``three-loop'' phase-space diagrams.
After performing the relevant Dirac algebra, each of these diagrams can be written as a 
linear combination of ``three-loop''
phase space integrals. We need to organize these integrals  
into \emph{integral families}, in order to perform a reduction to master
integrals using the well-known integration-by-parts identities~\cite{Chetyrkin:1981qh}.
When dealing with phase space integrals subject to  constraints, 
this step involves one more subtlety compared to 
what is done for standard multi-loop Feynman integrals.
In order to understand this, we recall that
a complete integral family should 
contain \emph{exactly} as many propagators, as is the total number
of independent scalar products that can be formed from  the loop momenta and the
external momenta. In our case, there are effectively three loop 
momenta and two external
momenta ($p$ and $\bar{p}$); it follows that   
we can construct 12 independent scalar products. Therefore, we need 
to map our phase-space integrals to integral families with 12 independent propagators.

This requires some extra work. 
Indeed, consider again \cref{fig:1}. 
It is easy to convince oneself that, after Dirac algebra, the phase-space integrals stemming from these diagrams will have the following general form
\begin{align}
\mathcal{I} \sim
\int \frac{d^d k_1}{(2\pi)^{d-1}}
\frac{d^d k_2}{(2\pi)^{d-1}}
\frac{d^d k_3}{(2\pi)^{d}}
\frac{
\delta^{+\!\!}\left(k_1^2\right)
\delta^{+\!\!}\left(k_2^2\right)
\delta\left(p \cdot k_{12} - \tfrac{t}{2z}\right)
\delta\left(\pbar \cdot k_{12} - \tfrac{1-z}{2} s \right)\, \mathcal{N}
}{
D_1^{n_1}\,
D_2^{n_2}\,
D_3^{n_3}\,
D_4^{n_4}\,
D_5^{n_5}\,
D_6^{n_6}\,
D_7^{n_7}\,
D_8^{n_8}\,
D_9^{n_9}\,
D_{10}^{n_{10}}\,
D_{11}^{n_{11}}\,
}\,.
\label{eq:topoI1}
\end{align}
Here $k_1,k_2$ are the momenta of two on-shell gluons and $k_3$ is 
the momentum of the  virtual gluon. 
The $D_j$ are different, unspecified, denominators and
$\mathcal{N}$ is a polynomial of  scalar products of the loop momenta and
the external momenta $p$, $\bar{p}$.
It should be easy to see that the diagrams obtained from \cref{fig:1} can generate at most 11 different propagators, including the combinations $k_{1} \cdot \bar{p}$ and $k_{2} \cdot \bar{p}$ which come from the sum over polarizations in \cref{eq:pol}, along with similar denominator factors coming from the gluon propagators in axial gauge.
Counting four delta-functions as ``propagators'' (we have in mind the reverse unitarity relation $\delta(x) \stackrel{\sim}{\to} 1/x$), we get 15 propagators in total. 
Since this number is larger than 12, the number of independent scalar products involving the momenta $k_i$, the propagators are linearly dependent and \cref{eq:topoI1} therefore does not constitute an integral family.

To remedy this problem, for every contributing diagram we partial fraction some of the linearly dependent propagators. 
For instance, we write 
\begin{align}
&\frac{1}{(k_1 \cdot \bar{p}) (k_2 \cdot \bar{p})} = \frac{2}{s(1-z)} \left[ \frac{1}{k_1 \cdot \bar{p}} + \frac{1}{k_2 \cdot \bar{p}}\right], \\
&\frac{1}{(k_1+k_2)^2 (k_1+k_2-p)^2} = \frac{z}{t} \left[ \frac{1}{(k_1+k_2-p)^2} - \frac{1}{(k_1+k_2)^2} \right], \\
&\frac{1}{(k_1-p)^2(k_2-p)^2} = -\frac{z}{t} \left[ \frac{1}{(k_1-p)^2} + \frac{1}{(k_2-p)^2}\right].
\end{align}
The partial fractioning effectively splits the diagrams into several terms, each of which contains at most 12 linearly independent propagators. 
At that point we can introduce well-defined integral families.

With this procedure we find that all diagrams can be expressed in terms of 19 independent integral families. 
The corresponding integrals can be written as 
\begin{align}
\mathcal{I}^{\text{top}}_{n_1, \dotsc, n_{8}} =
\int \frac{d^d k_1}{(2\pi)^{d-1}}
\frac{d^d k_2}{(2\pi)^{d-1}}
\frac{d^d k_3}{(2\pi)^{d}}
\frac{
\delta^{+\!\!}\left(k_1^2\right)
\delta^{+\!\!}\left(k_2^2\right)
\delta\left(p \cdot k_{12} - \tfrac{t}{2z}\right)
\delta\left(\pbar \cdot k_{12} - \tfrac{1-z}{2} s \right)
}{
D_1^{n_1}\,
D_2^{n_2}\,
D_3^{n_3}\,
D_4^{n_4}\,
D_5^{n_5}\,
D_6^{n_6}\,
D_7^{n_7}\,
D_8^{n_8}
}\,,
\label{eq:topoI}
\end{align}
where $\text{top} \in \{\text{A1},\text{A2},\dotsc,\text{A19}\}$ and the inverse propagators $D_{1}, \dotsc, D_{8}$ for each 
integral family are shown in \cref{tab:inverseprops}.

\begin{table}[t]
\begin{tabular}[c]{l|llllllll}
\text{top} ~~&~
$D_1$ & 
$D_2$ & 
$D_3$ & 
$D_4$ & 
$D_5$ & 
$D_{6}$ & 
$D_{7}$ & 
$D_{8}$ \\
\hline
 \text{A1} & ~~$k_3^2$ & $k_{12}^2$ & $k_{13}^2$ & $k_{123}^2$ & $(p-k_1)^2$ & $(p-k_{123})^2$ & $\pbar\cdot k_1$ & $\pbar\cdot k_3 $\\
 \text{A2} & ~~$k_3^2$ & $k_{13}^2$ & $k_{123}^2$ & $(p-k_1)^2$ & $(p-k_{12})^2$ & $(p-k_{123})^2$ & $\pbar\cdot k_1$ & $\pbar\cdot k_3 $\\
 \text{A3} & ~~$k_3^2$ & $k_{12}^2$ & $k_{13}^2$ & $k_{123}^2$ & $(p-k_1)^2$ & $(p-k_{123})^2$ & $\pbar\cdot k_2$ & $\pbar\cdot k_3 $\\
 \text{A4} & ~~$k_3^2$ & $k_{13}^2$ & $k_{123}^2$ & $(p-k_1)^2$ & $(p-k_{12})^2$ & $(p-k_{123})^2$ & $\pbar\cdot k_2$ & $\pbar\cdot k_3 $\\
 \text{A5} & ~~$k_3^2$ & $k_{23}^2$ & $(p-k_1)^2$ & $(p-k_{12})^2$ & $(p-k_{23})^2$ & $(p-k_{123})^2$ & $\pbar\cdot k_1$ & $\pbar\cdot k_3 $\\
 \text{A6} & ~~$k_3^2$ & $k_{13}^2$ & $k_{123}^2$ & $(p-k_1)^2$ & $(p-k_{12})^2$ & $(p-k_{123})^2$ & $\pbar\cdot k_1$ & $\pbar\cdot k_{123} $\\
 \text{A7} & ~~$k_3^2$ & $k_{12}^2$ & $k_{13}^2$ & $k_{123}^2$ & $(p-k_1)^2$ & $(p-k_{13})^2$ & $\pbar\cdot k_1$ & $\pbar\cdot k_3 $\\
 \text{A8} & ~~$k_3^2$ & $k_{12}^2$ & $k_{13}^2$ & $k_{123}^2$ & $(p-k_1)^2$ & $(p-k_{123})^2$ & $\pbar\cdot k_1$ & $\pbar\cdot k_{123} $\\
 \text{A9} & ~~$k_3^2$ & $k_{12}^2$ & $(p-k_1)^2$ & $(p-k_3)^2$ & $(p-k_{13})^2$ & $(p-k_{123})^2$ & $\pbar\cdot k_1$ & $\pbar\cdot k_3 $\\
 \text{A10} & ~~$k_3^2$ & $k_{13}^2$ & $(p-k_1)^2$ & $(p-k_{12})^2$ & $(p-k_{13})^2$ & $(p-k_{123})^2$ & $\pbar\cdot k_1$ & $\pbar\cdot k_3 $\\
 \text{A11} & ~~$k_3^2$ & $k_{12}^2$ & $(p-k_1)^2$ & $(p-k_3)^2$ & $(p-k_{13})^2$ & $(p-k_{123})^2$ & $\pbar\cdot k_2$ & $\pbar\cdot k_3 $\\
 \text{A12} & ~~$k_3^2$ & $(p-k_1)^2$ & $(p-k_3)^2$ & $(p-k_{12})^2$ & $(p-k_{13})^2$ & $(p-k_{123})^2$ & $\pbar\cdot k_1$ & $\pbar\cdot k_3 $\\
 \text{A13} & ~~$k_3^2$ & $(p-k_1)^2$ & $(p-k_3)^2$ & $(p-k_{12})^2$ & $(p-k_{13})^2$ & $(p-k_{123})^2$ & $\pbar\cdot k_2$ & $\pbar\cdot k_3 $\\
 \text{A14} & ~~$k_3^2$ & $k_{13}^2$ & $(p-k_1)^2$ & $(p-k_{12})^2$ & $(p-k_{13})^2$ & $(p-k_{123})^2$ & $\pbar\cdot k_1$ & $\pbar\cdot k_{13} $\\
 \text{A15} & ~~$k_3^2$ & $(p-k_1)^2$ & $(p-k_3)^2$ & $(p-k_{12})^2$ & $(p-k_{23})^2$ & $(p-k_{123})^2$ & $\pbar\cdot k_1$ & $\pbar\cdot k_3 $\\
 \text{A16} & ~~$k_3^2$ & $(p-k_1)^2$ & $(p-k_3)^2$ & $(p-k_{12})^2$ & $(p-k_{23})^2$ & $(p-k_{123})^2$ & $\pbar\cdot k_2$ & $\pbar\cdot k_3 $\\
 \text{A17} & ~~$k_3^2$ & $k_{12}^2$ & $k_{23}^2$ & $k_{123}^2$ & $(p-k_1)^2$ & $(p-k_{123})^2$ & $\pbar\cdot k_1$ & $\pbar\cdot k_3 $\\
 \text{A18} & ~~$k_3^2$ & $k_{23}^2$ & $k_{123}^2$ & $(p-k_1)^2$ & $(p-k_{12})^2$ & $(p-k_{123})^2$ & $\pbar\cdot k_1$ & $\pbar\cdot k_3 $\\
 \text{A19} & ~~$k_3^2$ & $k_{23}^2$ & $k_{123}^2$ & $(p-k_1)^2$ & $(p-k_{12})^2$ & $(p-k_{123})^2$ & $\pbar\cdot k_1$ & $\pbar\cdot k_{123} $
\end{tabular}
\caption{The inverse propagators $D_i$ for each of the 19 topologies A1, $\dotsc$, A19. Here we use the shorthand notation $k_{ij} = k_{i} + k_{j}$ and $k_{ij\ell} = k_{i} + k_{j} + k_{\ell}$.
\vspace{5mm}
}
\label{tab:inverseprops}
\end{table}

For each integral family we perform the reduction to master integrals using Reduze 2~\cite{vonManteuffel:2012np}, which supports operations with cut propagators. 
Performing the reduction, we find that all the contributing integrals can be expressed through 128 master integrals. 
We choose the set of master integrals as listed in \cref{tab:listofmasters}.

\begin{table}[t]
\centering
\begin{tabular}[c]{llllll}
$\mathcal{I}^{\text{A1}}_{0,0,1,0,0,1,0,0}$ & 
$\mathcal{I}^{\text{A1}}_{1,0,0,0,0,1,0,0}$ & 
$\mathcal{I}^{\text{A1}}_{1,0,0,1,0,0,0,0}$ &
$\mathcal{I}^{\text{A1}}_{1,-1,0,0,0,1,0,0}$ & 
$\mathcal{I}^{\text{A1}}_{0,0,1,0,0,1,0,1}$ &
$\mathcal{I}^{\text{A1}}_{1,0,1,0,0,1,0,0}$ 
\\[0.5mm]
$\mathcal{I}^{\text{A1}}_{-1,0,1,0,0,1,0,1}$ & 
$\mathcal{I}^{\text{A1}}_{1,-1,1,0,0,1,0,0}$ & 
$\mathcal{I}^{\text{A1}}_{0,1,1,0,0,1,0,1}$ & 
$\mathcal{I}^{\text{A1}}_{0,1,1,0,0,1,1,0}$ &
$\mathcal{I}^{\text{A1}}_{1,0,0,0,1,1,1,0}$ & 
$\mathcal{I}^{\text{A1}}_{1,0,0,1,0,1,0,1}$ 
\\[0.5mm]
$\mathcal{I}^{\text{A1}}_{1,0,1,0,0,1,0,1}$ & 
$\mathcal{I}^{\text{A1}}_{1,0,1,0,0,1,1,0}$ & 
$\mathcal{I}^{\text{A1}}_{-1,1,1,0,0,1,0,1}$ &
$\mathcal{I}^{\text{A1}}_{1,0,1,0,1,1,0,1}$ & 
$\mathcal{I}^{\text{A1}}_{1,0,1,1,0,1,1,0}$ & 
$\mathcal{I}^{\text{A1}}_{1,0,1,1,1,0,0,1}$ 
\\[0.5mm]
$\mathcal{I}^{\text{A1}}_{1,0,1,1,1,1,0,0}$ & 
$\mathcal{I}^{\text{A1}}_{1,-1,1,0,1,1,0,1}$ &
$\mathcal{I}^{\text{A1}}_{1,0,1,-1,1,1,0,1}$ & 
$\mathcal{I}^{\text{A1}}_{0,1,1,1,0,1,1,1}$ & 
$\mathcal{I}^{\text{A2}}_{0,1,0,0,1,1,0,0}$ & 
$\mathcal{I}^{\text{A2}}_{1,0,1,0,1,0,0,0}$ 
\\[0.5mm]
$\mathcal{I}^{\text{A2}}_{0,1,0,0,1,1,0,1}$ &
$\mathcal{I}^{\text{A2}}_{0,1,0,0,1,1,1,0}$ & 
$\mathcal{I}^{\text{A2}}_{-1,1,0,0,1,1,0,1}$ & 
$\mathcal{I}^{\text{A2}}_{0,1,-1,0,1,1,0,1}$ & 
$\mathcal{I}^{\text{A2}}_{0,1,0,1,1,1,0,1}$ & 
$\mathcal{I}^{\text{A2}}_{0,1,0,1,1,1,1,0} $
\\[0.5mm]
$\mathcal{I}^{\text{A2}}_{0,1,1,0,1,1,0,1}$ & 
$\mathcal{I}^{\text{A2}}_{1,0,1,0,1,1,0,1}$ & 
$\mathcal{I}^{\text{A2}}_{0,1,0,1,1,1,-1,1}$ & 
$\mathcal{I}^{\text{A2}}_{0,1,1,0,1,1,1,1}$ & 
$\mathcal{I}^{\text{A2}}_{1,1,1,0,1,1,1,0} $ &
$\mathcal{I}^{\text{A2}}_{1,1,1,1,1,0,0,1}$ 
\\[0.5mm]
$\mathcal{I}^{\text{A2}}_{1,1,1,1,1,1,0,0}$ & 
$\mathcal{I}^{\text{A2}}_{1,1,1,-1,1,1,1,0}$ & 
$\mathcal{I}^{\text{A2}}_{1,1,1,1,1,-1,0,1}$ & 
$\mathcal{I}^{\text{A3}}_{1,0,0,0,1,1,1,0} $ &
$\mathcal{I}^{\text{A3}}_{1,0,0,1,1,0,1,0}$ & 
$\mathcal{I}^{\text{A3}}_{1,0,1,0,0,1,1,0}$ 
\\[0.5mm]
 $\mathcal{I}^{\text{A3}}_{1,0,1,0,0,1,1,1}$ & 
 $\mathcal{I}^{\text{A3}}_{1,0,1,0,1,1,1,0}$ & 
 $\mathcal{I}^{\text{A3}}_{1,0,1,1,0,0,1,1} $ &
 $\mathcal{I}^{\text{A3}}_{1,0,1,1,0,1,1,0}$ & 
 $\mathcal{I}^{\text{A3}}_{1,1,0,0,1,1,1,0}$ & 
 $\mathcal{I}^{\text{A3}}_{1,0,1,0,1,1,1,1}$ 
\\[0.5mm]
 $\mathcal{I}^{\text{A3}}_{1,1,0,1,1,1,1,1}$ & 
 $\mathcal{I}^{\text{A4}}_{0,1,0,0,1,1,1,0} $ &
 $\mathcal{I}^{\text{A4}}_{0,1,0,1,1,1,1,0}$ & 
 $\mathcal{I}^{\text{A4}}_{1,0,1,1,1,0,1,0}$ & 
 $\mathcal{I}^{\text{A4}}_{1,1,1,0,1,0,1,1}$ & 
 $\mathcal{I}^{\text{A4}}_{1,1,1,0,1,1,1,0}$ 
\\[0.5mm]
 $\mathcal{I}^{\text{A4}}_{1,1,1,-1,1,1,1,0}$ &
 $\mathcal{I}^{\text{A5}}_{1,0,0,0,1,0,0,1}$ & 
 $\mathcal{I}^{\text{A5}}_{0,1,0,0,1,1,0,1}$ & 
 $\mathcal{I}^{\text{A5}}_{1,0,0,0,1,1,0,1}$ & 
 $\mathcal{I}^{\text{A5}}_{1,0,0,1,1,0,0,1}$ & 
 $\mathcal{I}^{\text{A5}}_{1,1,0,0,1,0,0,1} $
\\[0.5mm]
 $\mathcal{I}^{\text{A5}}_{0,1,0,1,1,1,0,1}$ & 
 $\mathcal{I}^{\text{A5}}_{1,0,0,0,1,1,1,1}$ & 
 $\mathcal{I}^{\text{A5}}_{1,0,0,1,1,1,0,1}$ & 
 $\mathcal{I}^{\text{A5}}_{1,0,1,0,1,1,0,1}$ & 
 $\mathcal{I}^{\text{A5}}_{1,0,1,1,1,0,0,1}$ &
 $\mathcal{I}^{\text{A5}}_{1,1,0,0,1,1,0,1}$
\\[0.5mm]
 $\mathcal{I}^{\text{A5}}_{1,1,0,0,1,1,1,0}$ & 
 $\mathcal{I}^{\text{A5}}_{1,1,0,1,1,0,0,1}$ & 
 $\mathcal{I}^{\text{A5}}_{1,1,0,1,1,1,0,0}$ & 
 $\mathcal{I}^{\text{A5}}_{1,1,1,0,0,1,0,1}$ &
 $\mathcal{I}^{\text{A5}}_{-1,1,0,1,1,1,0,1}$ &
 $\mathcal{I}^{\text{A5}}_{0,1,-1,1,1,1,0,1}$ 
\\[0.5mm]
 $\mathcal{I}^{\text{A5}}_{1,-1,1,0,1,1,0,1}$ & 
 $\mathcal{I}^{\text{A5}}_{1,1,-1,1,1,0,0,1}$ & 
 $\mathcal{I}^{\text{A5}}_{1,0,1,1,1,1,0,1} $ &
 $\mathcal{I}^{\text{A5}}_{1,1,0,1,1,0,1,1}$ & 
 $\mathcal{I}^{\text{A5}}_{1,1,0,1,1,1,1,0}$ & 
 $\mathcal{I}^{\text{A5}}_{1,1,1,0,0,1,1,1}$ 
\\[0.5mm]
 $\mathcal{I}^{\text{A5}}_{1,1,1,1,1,0,0,1}$ & 
 $\mathcal{I}^{\text{A5}}_{1,1,-1,1,1,1,1,0}$ &
 $\mathcal{I}^{\text{A5}}_{1,1,1,1,1,0,1,1}$ & 
 $\mathcal{I}^{\text{A6}}_{1,1,1,0,1,0,1,1}$ & 
 $\mathcal{I}^{\text{A6}}_{1,1,1,1,1,0,0,1}$ & 
 $\mathcal{I}^{\text{A6}}_{1,1,1,1,1,-1,0,1}$ 
\\[0.5mm]
 $\mathcal{I}^{\text{A7}}_{1,1,0,0,0,1,0,1} $ &
 $\mathcal{I}^{\text{A8}}_{1,0,1,1,0,0,1,1}$ & 
 $\mathcal{I}^{\text{A8}}_{1,0,1,1,1,0,0,1}$ & 
 $\mathcal{I}^{\text{A9}}_{0,0,0,1,0,1,0,1}$ & 
 $\mathcal{I}^{\text{A9}}_{1,0,0,1,0,1,0,1}$ & 
 $\mathcal{I}^{\text{A9}}_{0,0,0,1,1,1,1,1} $
\\[0.5mm]
 $\mathcal{I}^{\text{A9}}_{0,0,1,1,1,1,0,1}$ & 
 $\mathcal{I}^{\text{A9}}_{1,0,0,1,1,0,1,1}$ & 
 $\mathcal{I}^{\text{A9}}_{1,0,1,0,1,1,0,1}$ & 
 $\mathcal{I}^{\text{A9}}_{-1,0,1,1,1,1,0,1}$ & 
 $\mathcal{I}^{\text{A9}}_{1,0,0,1,1,1,1,1} $ &
 $\mathcal{I}^{\text{A9}}_{1,0,1,0,1,1,1,1}$ 
\\[0.5mm]
 $\mathcal{I}^{\text{A9}}_{1,1,1,0,1,1,0,1}$ & 
$\mathcal{I}^{\text{A9}}_{1,1,1,-1,1,1,0,1}$ & 
$\mathcal{I}^{\text{A10}}_{0,1,0,1,1,1,1,1}$ & 
$\mathcal{I}^{\text{A10}}_{0,1,1,1,1,1,0,1} $ &
$\mathcal{I}^{\text{A10}}_{0,1,1,1,1,1,1,1}$ & 
$\mathcal{I}^{\text{A11}}_{0,0,0,1,1,1,1,1}$ 
\\[0.5mm]
$\mathcal{I}^{\text{A11}}_{0,0,1,1,0,1,1,1}$ & 
$\mathcal{I}^{\text{A11}}_{1,0,0,1,1,1,1,1}$ & 
$\mathcal{I}^{\text{A11}}_{1,0,1,1,0,1,1,1} $ &
$\mathcal{I}^{\text{A11}}_{1,1,0,1,1,0,1,1}$ & 
$\mathcal{I}^{\text{A11}}_{1,1,1,1,0,1,1,1}$ & 
$\mathcal{I}^{\text{A12}}_{0,0,1,1,0,1,0,1}$ 
\\[0.5mm]
$\mathcal{I}^{\text{A12}}_{1,0,1,1,1,0,0,1}$ & 
$\mathcal{I}^{\text{A12}}_{0,1,1,1,1,1,0,1} $ &
$\mathcal{I}^{\text{A12}}_{-1,1,1,1,1,1,0,1}$ & 
$\mathcal{I}^{\text{A13}}_{1,0,1,1,1,0,1,1}$ & 
$\mathcal{I}^{\text{A14}}_{0,1,1,1,1,1,0,1}$ & 
$\mathcal{I}^{\text{A14}}_{0,1,1,1,1,1,1,1}$ 
\\[0.5mm]
$\mathcal{I}^{\text{A15}}_{0,1,1,0,1,1,0,1} $ &
$\mathcal{I}^{\text{A15}}_{-1,1,1,0,1,1,0,1}$ & 
$\mathcal{I}^{\text{A15}}_{0,1,1,1,1,1,0,1}$ & 
$\mathcal{I}^{\text{A15}}_{1,1,1,0,1,1,0,1}$ & 
$\mathcal{I}^{\text{A15}}_{-1,1,1,1,1,1,0,1}$ & 
$\mathcal{I}^{\text{A16}}_{1,1,1,0,1,1,1,1} $ 
\\[0.5mm]
$\mathcal{I}^{\text{A17}}_{1,0,1,1,1,0,0,1}$ & 
$\mathcal{I}^{\text{A17}}_{1,1,1,0,1,1,0,1}$ & 
$\mathcal{I}^{\text{A17}}_{1,1,1,-1,1,1,0,1}$ & 
$\mathcal{I}^{\text{A18}}_{1,1,1,1,1,0,0,1}$ & 
$\mathcal{I}^{\text{A18}}_{1,1,1,1,1,-1,0,1}$ &
$\mathcal{I}^{\text{A19}}_{1,1,1,1,0,0,0,1}$ 
\\[0.5mm]
$\mathcal{I}^{\text{A19}}_{1,1,1,1,1,0,0,1}$ & 
$\mathcal{I}^{\text{A19}}_{1,1,1,1,1,-1,0,1}$
\end{tabular}
\caption{The set of 128 master integrals for the matching coefficients of the double-real contribution to the quark beam function.}
\label{tab:listofmasters}
\end{table}
In the next section we discuss in detail how to evaluate these 128 master integrals using the method of 
differential equations~\cite{Kotikov:1990kg,Bern:1993kr,Remiddi:1997ny,Gehrmann:1999as} augmented by the use of a canonical basis~\cite{Henn:2013pwa}.


\section{Master integrals}
\label{sec:masters}

Having expressed the amplitude in terms of master integrals, we proceed with the discussion of their evaluation. 
The master integrals depend on the three quantities $s = 2 p \cdot \pbar$, $\,t$ and $z$, cf. \cref{eq:topoI}. However, some of these dependencies are quite simple. 
Indeed, we re-define the gluon momenta as 
$k_i = {\tilde k}_i \, \sqrt{t/z},~ 
p = {\tilde p} \, \sqrt{t/z}$ and $\pbar = {\tilde \pbar} \, s\sqrt{z/t}$.
Since $p \cdot \bar p  =s/2$, we find ${\tilde p} \cdot {\tilde \pbar} =1/2$.
Applying these transformations to integrands  of master integrals, we find 
\begin{align}
\mathcal{I}^{\text{top}}_{n_1, \dotsc, n_{8}}(s,t,z) &= \left(\frac{1}{s}\right)^{1+n_7+n_8} 
\left(\frac{t}{z}\right)^{3 - (n_1+n_2+\dotsc+n_6) - 3 \eps} 
\mathcal{I}^{\text{top}}_{n_1, \dotsc, n_{8}}(1,z,z)\,, 
\label{eq:rescaling-step1}
\\
&\equiv \left(\frac{1}{s}\right)^{1+n_7+n_8} \left(\frac{t}{z}\right)^{3 - (n_1+n_2+\dotsc+n_6) - 3 \eps} \mathcal{I}^{\text{top}}_{n_1, \dotsc, n_{8}}(z)\,.
\label{eq:rescaling}
\end{align}
Note that the second step in the above equation implies that one can get the 
expression for the re-scaled integral upon taking the definition of the original 
integral with its full $s,t$ and $z$ dependence and then evaluating it for $s=1$ and  
$t=z$. 

In order to calculate the master integrals $\mathcal{I}^{\text{top}}_{n_1, \dotsc, n_{8}}(z)$ 
on the right-hand side of \cref{eq:rescaling}, we compute 
their derivative with respect to $z$ and re-express 
the result in terms of master integrals by means of 
integration-by-parts reduction, thus producing a closed system of differential equations.
We write it as 
\begin{align}
\frac{d}{dz} \vec{\mathcal{I}}(z,\eps) = \hat A(z,\eps)\, \vec{\mathcal{I}}(z,\eps)\,,
\end{align}
where $\vec{\mathcal{I}}(z,\eps)$ contains the $128$ master integrals in 
\cref{tab:listofmasters} with $s=1$ and $t=z$, and $\hat A(z,\eps)$ is a $128 \times 128$ matrix.
We use the program \texttt{Fuchsia} \cite{Gituliar:2017vzm} 
to transform the system of equation to the  $\epsilon$-form 
using the algorithm described in ref.~\cite{Lee:2014ioa}. We find 
\begin{align}
\frac{d}{d z} \vec{\mathcal{I}}_{\text{can}}(z,\eps) = \eps \,
\left( 
\frac{\hat A_{-1}}{z+1} + 
\frac{\hat A_0}{z} + 
\frac{\hat A_1}{z-1} + 
\frac{\hat A_2}{z-2}
\right) 
\vec{\mathcal{I}}_{\text{can}}(z,\eps)\,,
\label{eq:deq}
\end{align}
where the $\hat A_k$ are constant matrices.  The vector  $\vec{\mathcal{I}}_{\text{can}}(z,\eps)$ contains 
the canonical master integrals; it is related to the original master integrals
$\vec{\mathcal{I}}(z,\eps) = \hat T(z,\eps)\,\vec{\mathcal{I}}_{\text{can}}(z,\eps)$ 
by a transformation matrix $\hat T(z,\eps)$.

The system of differential equations in \cref{eq:deq} is solved iteratively for the coefficients 
of $\vec{\mathcal{I}}_{\text{can}}(z,\eps)$ in an expansion around $\eps=0$.
The result of this procedure can be written as
\begin{align}
\vec{\mathcal{I}}_{\text{can}}(z,\eps) = \hat M(z,\eps)\, \vec{B}(\eps)\,,
\label{eq:sol}
\end{align}
where the $128 \times 128$ matrix $\hat M(z,\eps)$ contains elements of the form
\begin{align}
M_{ij}(z,\eps) = \sum_{k=0}^{6} \, \sum_{\vec{w} \,\in\, W(k)} c_{i,j,k,\vec{w}} \,\, \eps^k \, G(\vec{w}; z)\,.
\end{align}
The inner sum runs over vectors $\vec{w}$ of length $k$ whose 
components are drawn from a set $\{-1,0,1,2\}$, 
which are the singular points in the differential equations \cref{eq:deq}.
The $G(\vec{w}; z)$ are multiple polylogarithms~\cite{Goncharov:1998kja,Remiddi:1999ew,Goncharov:2001iea,Vollinga:2004sn}, which are defined iteratively as
\begin{align}
G(w_1,w_2,\dotsc,w_n; z) = \int_0^z  dt \,\frac{G(w_2,w_3,\dotsc,w_n ; t)}{t-w_1}\,,
\label{eq:gpl}
\end{align}
with the special cases
\begin{align}
G(; z) = 1~,~~~ G(\underbrace{0,\dotsc,0}_{n \,\text{times}}; z) = \frac{1}{n!} \log^n(z)\,.
\label{eq:gpl-start}
\end{align}
The constant vector  $\vec{B}(\eps)$ in \cref{eq:sol} is 
fixed by the boundary conditions required 
for the solution of  the system of first-order differential equations.
We extract the boundary condition 
from the behaviour of the master integrals in the limit $z \to 1$.

In principle, it is possible to extract the boundary conditions by writing
\begin{align}
\vec{\mathcal{I}}(z,\eps) = \hat T(z,\eps)\,\hat M(z,\eps)\,\vec{B}(\eps)\,,
\label{eq:ItoB}
\end{align}
and taking the limit $z \to 1$. 
The multiple polylogarithms $G(\vec{w}; z)$ inside $\hat M(z,\eps)$, appearing on the right-hand side of \cref{eq:ItoB}, are, in general, 
logarithmically divergent in this limit. 
This divergence should be extracted by writing the multiple polylogarithms in the form $G(\vec{w}; z) = \sum_{\ell} c_\ell \log^\ell(1-z)$.
Computing a sufficient number of integrals $\vec{\mathcal{I}}(z,\eps)$ 
that appear on the left-hand side of \cref{eq:ItoB} 
in the $z \to 1$ limit  allows to determine the constants $\vec{B}(\eps)$.

In performing the computation of the master integrals we should remember that 
the beam function matching coefficients should be treated as
distributions in $(1-z)$.
Indeed, the matching coefficients  have to be convoluted with the PDFs and all divergences in the limit $z \to 1$ have to be regularized, 
extracted and renormalized in order to obtain the final finite result in terms of quantities like 
$\delta(1-z)$, $[1/(1-z)]_+$, etc.
To extract these distributions, one needs to have the master integrals written in  the 
``resummed'' form, i.e. the $\epsilon$ dependence in the limit $z \to 1$ has to be made explicit
in the form of $(1-z)^{a+b\epsilon}$ powers.  It is convenient, therefore, to extract 
the boundary constants in a similar way, namely by matching 
equal powers of $(1-z)^{\eps}$ on both sides of \cref{eq:ItoB}.

In order to do that, the right-hand side of \cref{eq:ItoB} should first be 
written in the  ``resummed'' form $\sum_{\ell,m} c_{\ell,m}(\eps) (1-z)^{n + \ell \eps} \log^m(1-z)$, 
which can be easily achieved by solving the differential equations in \cref{eq:deq} in the limit $z \to 1$
\begin{align}
  \frac{d}{d z} \vec{\mathcal{I}}_{\text{can}}(z \to 1,\eps) = \eps \, \frac{\hat A_1}{z-1}\,
  \vec{\mathcal{I}}_{\text{can}}(z \to 1,\eps)\,,
\label{eq:limit-deq}
\end{align}
whose solution is given by a matrix exponential 
\begin{align}
\vec{\mathcal{I}}_{\text{can}}(z \to 1,\eps) = (1-z)^{- \eps \hat A_1} \vec{H}(\eps)\,.
\label{eq:limit-sol}
\end{align}
The new constants in $\vec{H}(\eps)$ can be expressed as linear combinations of the constants in $\vec{B}(\eps)$ 
by equating the right-hand side of \cref{eq:sol} taken in the limit $z \to 1$ and the right-hand side of \cref{eq:limit-sol} expanded in $\eps$.
Multiplying both sides in \cref{eq:limit-sol} by the transformation matrix $\hat T(z,\eps)$ in the limit $z \to 1$ and performing the matrix exponentiation gives the leading-power behaviour of the master integrals
\begin{align}
\mathcal{I}^{(i)}(z \to 1,\eps) = \sum_{\ell=-3}^{2} \sum_{m=0}^{2} C_{i,\ell,m}(\eps)\, (1-z)^{n(i)+\ell \eps} \log^m(1-z)\,,
~~(i=1,\dotsc,128;~n(i) \in \mathbb{Z})\,.
\label{eq:BehaviourI}
\end{align}
The coefficients $C_{i,\ell,m}(\eps)$ are in general linear combinations of the $128$ constants in $\vec{B}(\eps)$.
Therefore, we need to compute sufficiently many linearly independent $C_{i,\ell,m}(\eps)$ to determine all boundary constants.
Fortunately, we can fix many boundary constants by identifying the constants $C_{i,\ell,m}(\eps)$ that must vanish in order to produce the correct behaviour of the master integrals in the limit $z \to 1$. 

As an explicit example, consider the second master integral $\mathcal{I}^{(2)}(z \to 1) = \mathcal{I}^{\text{A1}}_{1,0,0,0,0,1,0,0}|_{z \to 1}$.
On the one hand, the differential equations predict that in the limit $z \to 1$ the integral has the form 
\begin{align}
\mathcal{I}^{(2)}(z \to 1) = C_{2,0,0}(\eps) + C_{2,-2,0}(\eps) (1-z)^{-2\eps} + \mathcal{O}(1-z)\,.
\end{align}
On the other hand, its integral representation suggests that its leading behaviour scales as $(1-z)^{1-2\eps}$, which implies that $C_{2,0,0}(\eps) = C_{2,-2,0}(\eps) = 0$.
Let us verify this by inspecting the integral representation
\begin{align}
\mathcal{I}^{(2)}(z \to 1) 
= 
\int \frac{d^d k_1}{(2\pi)^{d-1}}
\frac{d^d k_2}{(2\pi)^{d-1}}
\frac{d^d k_3}{(2\pi)^{d}}
\frac{
\delta^{+\!\!}\left(k_1^2\right)
\delta^{+\!\!}\left(k_2^2\right)
\delta\left(p \cdot k_{12} - \tfrac{1}{2}\right)
\delta\left(\pbar \cdot k_{12} - \tfrac{1-z}{2} \right)
}{k_3^2\,(p-k_{123})^2}\,.
\label{eq:Int3}
\end{align}
The integral over $k_3$ is straightforward to evaluate. The result is 
proportional to  ${[-(p-k_{12})^2]^{-\eps}}$ which becomes  $[1 - k_{12}^2]^{-\eps}$ 
upon imposing the delta-function $\delta(p\cdot k_{12} -\tfrac{1}{2})$ 
and the on-shellness $p^2=0$ constraints.  
The last delta-function in the numerator of \cref{eq:Int3} fixes the  
projection of the total emitted momentum $k_{12} \equiv k_{1}+k_{2}$ to be small in the limit $z\to 1$.
To capture that, it is convenient to make a Sudakov decomposition of $k_1$ and $k_2$
\begin{align}
k_i^\mu = \alpha_i \, p^\mu + \beta_i \, \pbar^\mu + k_{i \perp}^\mu\,.
\label{eq:Sudakov}
\end{align}
We use this decomposition to compute 
$2 \pbar \cdot k_{12} = \alpha_{12} \equiv \alpha_{1}+\alpha_{2} 
\sim \order{1-z}$ and $2 p \cdot k_{12} = \beta_{12} \equiv \beta_{1}+\beta_{2} \sim \order{1}$.
Moreover, the fact that the momenta $k_{1}$ and $k_{2}$ are both on-shell 
and have positive energy, implies that $\alpha_i \sim  \order{1-z}, \beta_{i} \sim  
\order{1}$ and $k_{i\perp}^2 \sim  \order{1-z}$ for $i=1,2$ separately.
As a consequence, $k_{12}^2 \sim  \order{1-z}$ and the result of the $k_3$ integral 
can be simplified  $[1 - k_{12}^2]^{-\eps} = 1 + \order{1-z}$.
The integration measures for $k_1$ and $k_2$ reads 
$d^d k_i = \frac{1}{4} \, d\alpha_i \, d\beta_i \, d(k_{i\perp}^2) \, (k_{i \perp}^2)^{-\eps} \, d\Omega_{i}^{(d-2)}$ and  
scale as $\order{(1-z)^{2-\eps}}$.
Three of the delta functions in  \cref{eq:Int3} scale as $\order{(1-z)^{-1}}$.
Putting everything together, we find 
that $\mathcal{I}^{(2)}(z \to 1) = \order{(1-z)^{1-2\eps}}$.
Therefore, there are no contributions to the integral that 
scale as $\order{(1-z)^{0}}$ or $\order{(1-z)^{-2\eps}}$. 
We conclude that the coefficients of these regions vanish: $C_{2,0,0}(\eps) = C_{2,-2,0}(\eps) = 0$.

After finding all the coefficients $C_{i,\ell,m}(\eps)$ that must vanish because of similar arguments, we acquire enough relations to express 104 of the boundary constants in $\vec{B}(\eps)$ in terms of a remaining set of 24 constants.
To determine the latter constants we performed explicit computations of non-vanishing regions of selected master integrals in the limit $z \to 1$.

The boundary integrals that we have calculated are
\begin{align}
\label{eq:regionB1}
B_{1} = \mathcal{I}_{0,0,1,0,0,1,0,0}^{\text{A1}}\big|_{s=1,t=z,z\approx 1}&= C_{1,-2,0}(\epsilon ) (1-z)^{1-2 \epsilon } ~,\\
B_{2} = \mathcal{I}_{1,0,0,1,0,0,0,0}^{\text{A1}}\big|_{s=1,t=z,z\approx 1}&= C_{3,-3,0}(\epsilon ) (1-z)^{1-3 \epsilon }  ~,\\
B_{3} = \mathcal{I}_{0,0,1,0,0,1,0,1}^{\text{A1}}\big|_{s=1,t=z,z\approx 1}&= C_{5,-3,0}(\epsilon ) (1-z)^{1-3 \epsilon } ~,\\
B_{4} = \mathcal{I}_{0,1,1,0,0,1,0,1}^{\text{A1}}\big|_{s=1,t=z,z\approx 1}&= C_{9,-3,0}(\epsilon ) (1-z)^{-3 \epsilon } ~,\\
B_{5} = \mathcal{I}_{0,1,1,0,0,1,1,0}^{\text{A1}}\big|_{s=1,t=z,z\approx 1}&= C_{10,-2,0}(\epsilon ) (1-z)^{ -1-2 \epsilon} ~,\\
B_{6} = \mathcal{I}_{1,0,0,1,0,1,0,1}^{\text{A1}}\big|_{s=1,t=z,z\approx 1}&= C_{12,-3,0}(\epsilon ) (1-z)^{-3 \epsilon } ~,\\
B_{7} = \mathcal{I}_{1,0,1,0,1,1,0,1}^{\text{A1}}\big|_{s=1,t=z,z\approx 1}&= C_{16,-3,0}(\epsilon ) (1-z)^{-3 \epsilon } ~,\\
B_{8} = \mathcal{I}_{1,0,1,1,0,1,1,0}^{\text{A1}}\big|_{s=1,t=z,z\approx 1}&= C_{17,-3,0}(\epsilon ) (1-z)^{ -1-3 \epsilon} ~,\\
B_{9} = \mathcal{I}_{1,0,1,1,1,0,0,1}^{\text{A1}}\big|_{s=1,t=z,z\approx 1}&= C_{18,-3,0}(\epsilon ) (1-z)^{ -1-3 \epsilon} ~,\\
B_{10} = \mathcal{I}_{1,0,1,1,1,1,0,0}^{\text{A1}}\big|_{s=1,t=z,z\approx 1}&= C_{19,-3,0}(\epsilon ) (1-z)^{-3 \epsilon } ~,\\
B_{11} = \mathcal{I}_{0,1,1,1,0,1,1,1}^{\text{A1}}\big|_{s=1,t=z,z\approx 1}&= C_{22,-3,0}(\epsilon ) (1-z)^{ -2-3 \epsilon} ~,\\
B_{12} = \mathcal{I}_{1,0,0,1,1,0,1,0}^{\text{A3}}\big|_{s=1,t=z,z\approx 1}&= C_{41,-3,0}(\epsilon ) (1-z)^{-3 \epsilon } ~,\\
B_{13} = \mathcal{I}_{1,0,1,0,1,1,1,0}^{\text{A3}}\big|_{s=1,t=z,z\approx 1}&= C_{44,-2,0}(\epsilon ) (1-z)^{-2 \epsilon } ~,\\
B_{14} = \mathcal{I}_{1,0,1,1,0,0,1,1}^{\text{A3}}\big|_{s=1,t=z,z\approx 1}&= C_{45,-3,0}(\epsilon ) (1-z)^{ -2-3 \epsilon} ~,\\
B_{15} = \mathcal{I}_{1,0,1,1,0,1,1,0}^{\text{A3}}\big|_{s=1,t=z,z\approx 1}&= C_{46,-3,0}(\epsilon ) (1-z)^{ -1-3 \epsilon} ~,\\
B_{16} = \mathcal{I}_{1,1,0,0,1,1,1,0}^{\text{A3}}\big|_{s=1,t=z,z\approx 1}&= C_{47,-2,0}(\epsilon ) (1-z)^{ -1-2 \epsilon} ~,\\
B_{17} = \mathcal{I}_{1,1,0,1,1,1,1,1}^{\text{A3}}\big|_{s=1,t=z,z\approx 1}&= C_{49,-3,0}(\epsilon ) (1-z)^{ -2-3 \epsilon} ~,\\
B_{18} = \mathcal{I}_{1,1,0,0,1,0,0,1}^{\text{A5}}\big|_{s=1,t=z,z\approx 1}&= C_{60,-3,0}(\epsilon ) (1-z)^{-3 \epsilon } ~,\\
B_{19} = \mathcal{I}_{1,1,0,0,1,1,1,0}^{\text{A5}}\big|_{s=1,t=z,z\approx 1}&= C_{67,-2,0}(\epsilon ) (1-z)^{-2 \epsilon } ~,\\
B_{20} = \mathcal{I}_{1,0,1,1,0,0,1,1}^{\text{A8}}\big|_{s=1,t=z,z\approx 1}&= C_{86,-3,0}(\epsilon ) (1-z)^{ -2-3 \epsilon} ~,\\
B_{21} = \mathcal{I}_{1,0,1,1,1,0,0,1}^{\text{A8}}\big|_{s=1,t=z,z\approx 1}&= C_{87,-3,0}(\epsilon ) (1-z)^{ -1-3 \epsilon} ~,\\
B_{22} = \mathcal{I}_{1,0,1,1,1,0,0,1}^{\text{A17}}\big|_{s=1,t=z,z\approx 1}&= C_{121,-3,0}(\epsilon ) (1-z)^{ -1-3 \epsilon} ~,\\
B_{23} = \mathcal{I}_{1,1,1,0,1,1,0,1}^{\text{A17}}\big|_{s=1,t=z,z\approx 1}&= C_{122,-3,0}(\epsilon ) (1-z)^{ -1-3 \epsilon} ~,\\
B_{24} = \mathcal{I}_{1,1,1,1,0,0,0,1}^{\text{A19}}\big|_{s=1,t=z,z\approx 1}&= C_{126,-3,0}(\epsilon ) (1-z)^{-1-3 \epsilon} ~.
\label{eq:regionB24}
\end{align}
In the following, we provide the results for these  constants  and present a few examples that illustrate how they are   evaluated.

\subsection{Results for explicitly calculated coefficients}
Here we list the results in terms of Laurent series in $\eps$ for the constants $C_{i,j,k}(\epsilon)$ that appear in \crefrange{eq:regionB1}{eq:regionB24}. In the following sections we provide the reader with some examples of the computations that led to the expressions listed below.
For convenience we extract a common $\eps$-dependent pre-factor,
\begin{align}
C_{i,j,k}(\epsilon) = i \left(\frac{\Omega_{d-2}}{(2\pi)^{d-1}}\right)^3 \widetilde{C}_{i,j,k}(\epsilon)\,.
\end{align}
The results for the constants, up to weight six,  are
\begin{align}
 \widetilde{C}_{1,-2,0}(\epsilon ) &=  \tfrac{1}{(1-3 \epsilon ) (1-2 \epsilon )^2}\left(\tfrac{1}{16 \epsilon }
 -\tfrac{\pi ^2 \epsilon }{32}-\tfrac{5 \zeta _3 \epsilon ^2}{8}-\tfrac{\pi ^4 \epsilon ^3}{128}
 +\left(\tfrac{5 \pi ^2 \zeta _3}{16}-\tfrac{27 \zeta _5}{8}\right) \epsilon ^4 \right. \nn\\& \left. 
 +\left(\tfrac{25 \zeta _3^2}{8}-\tfrac{521 \pi ^6}{241920}\right) \epsilon ^5+\mathcal{O}\left(\epsilon ^6\right)\right), \\
 \widetilde{C}_{3,-3,0}(\epsilon ) &=  \tfrac{e^{i \pi  \epsilon }}{(1-3 \epsilon ) (1-2 \epsilon )^2}\left(\tfrac{1}{16 \epsilon }
 -\tfrac{\pi ^2 \epsilon }{32}-\tfrac{5 \zeta _3 \epsilon ^2}{8}
 -\tfrac{\pi ^4 \epsilon ^3}{128}+\left(\tfrac{5 \pi ^2 \zeta _3}{16}-\tfrac{27 \zeta _5}{8}\right) \epsilon ^4 \right. \nn\\& \left. 
 +\left(\tfrac{25 \zeta _3^2}{8}-\tfrac{521 \pi ^6}{241920}\right) \epsilon ^5+\mathcal{O}\left(\epsilon ^6\right)\right), \\
 \widetilde{C}_{5,-3,0}(\epsilon ) &=  \tfrac{e^{i \pi  \epsilon }}{(1-3 \epsilon )^2}\left(\tfrac{1}{8 \epsilon ^2}
 -\tfrac{\pi ^2}{24}-\tfrac{3 \zeta _3 \epsilon }{2}-\tfrac{13 \pi ^4 \epsilon ^2}{360}+\left(\tfrac{\pi ^2 \zeta _3}{2}
 -\tfrac{21 \zeta _5}{2}\right) \epsilon ^3 \right. \nn\\& \left.
 +\left(9 \zeta _3^2-\tfrac{59 \pi ^6}{3780}\right) \epsilon ^4  
 +\mathcal{O}\left(\epsilon ^5\right)\right), \\
 \widetilde{C}_{9,-3,0}(\epsilon ) &=  \tfrac{e^{i \pi  \epsilon }}{1-4 \epsilon }\left(-\tfrac{1}{8 \epsilon ^3}
 +\tfrac{\pi ^2}{16 \epsilon }+\tfrac{19 \zeta _3}{8}+\tfrac{41 \pi ^4 \epsilon }{576}
 +\left(\tfrac{423 \zeta _5}{16}-\tfrac{29 \pi ^2 \zeta _3}{24}\right) \epsilon ^2 \right. \nn\\& \left. 
 +\left(\tfrac{1273 \pi ^6}{24192}-23 \zeta _3^2\right) \epsilon ^3
 +\mathcal{O}\left(\epsilon ^4\right)\right), \\
 \widetilde{C}_{10,-2,0}(\epsilon ) &=  \tfrac{1}{1-2 \epsilon }\left(\tfrac{1}{8 \epsilon ^3}
 -\tfrac{\pi ^2}{12 \epsilon }-2 \zeta _3-\tfrac{53 \pi ^4 \epsilon }{1440}
 +\left(\tfrac{29 \pi ^2 \zeta _3}{24}-\tfrac{129 \zeta _5}{8}\right) \epsilon ^2 \right. \nn\\& \left.
 +\left(\tfrac{55 \zeta _3^2}{4}-\tfrac{341 \pi ^6}{20160}\right) \epsilon ^3
 +\mathcal{O}\left(\epsilon ^4\right)\right), \\
 \widetilde{C}_{12,-3,0}(\epsilon ) &=  \tfrac{e^{i \pi  \epsilon }}{1-2 \epsilon }\left(-\tfrac{\pi ^2}{24 \epsilon }
 -\tfrac{3 \zeta _3}{4}-\tfrac{19 \pi ^4 \epsilon }{1440}+\left(\tfrac{19 \pi ^2 \zeta _3}{24}
 -\tfrac{75 \zeta _5}{8}\right) \epsilon ^2 \right. \nn\\& \left.
 +\left(\tfrac{15 \zeta _3^2}{2}
 -\tfrac{547 \pi ^6}{60480}\right) \epsilon ^3+\mathcal{O}\left(\epsilon ^4\right)\right), \\
 \widetilde{C}_{16,-3,0}(\epsilon ) &=  e^{i \pi  \epsilon } \left(-\tfrac{1}{16 \epsilon ^4}
 +\tfrac{\zeta _3}{8 \epsilon }+\tfrac{\pi ^4}{960}+\left(\tfrac{\pi ^2 \zeta _3}{8}
 -\tfrac{9 \zeta _5}{16}\right) \epsilon +\left(\tfrac{7 \zeta _3^2}{4}
 +\tfrac{227 \pi ^6}{120960}\right) \epsilon ^2+\mathcal{O}\left(\epsilon ^3\right)\right), \\
 \widetilde{C}_{17,-3,0}(\epsilon ) &=  e^{i \pi  \epsilon } \left(-\tfrac{5}{16 \epsilon ^4}
 +\tfrac{11 \pi ^2}{32 \epsilon ^2}+\tfrac{11 \zeta _3}{\epsilon }
 +\tfrac{193 \pi ^4}{640}+\left(\tfrac{549 \zeta _5}{4}-\tfrac{17 \pi ^2 \zeta _3}{2}\right) \epsilon \right. \nn\\& \left. 
 +\left(\tfrac{47227 \pi ^6}{241920}-118 \zeta _3^2\right) \epsilon ^2+\mathcal{O}\left(\epsilon ^3\right)\right), \\
 \widetilde{C}_{18,-3,0}(\epsilon ) &=  \tfrac{e^{i \pi  \epsilon }}{1+4 \epsilon}\left(\tfrac{19}{32 \epsilon ^3}
 -\tfrac{19 \pi ^2}{64 \epsilon }-\tfrac{77 \zeta _3}{16}+\tfrac{49 \pi ^4 \epsilon }{1280}
 +\left(\tfrac{77 \pi ^2 \zeta _3}{32}+\tfrac{441 \zeta _5}{16}\right) \epsilon ^2 \right. \nn\\& \left.
 +\left(\tfrac{403 \zeta _3^2}{16}+\tfrac{21953 \pi ^6}{96768}\right) \epsilon ^3
 +\mathcal{O}\left(\epsilon ^4\right)\right), \\
 \widetilde{C}_{19,-3,0}(\epsilon ) &=  e^{i \pi  \epsilon } \left(\tfrac{3}{32 \epsilon ^4}-\tfrac{11 \pi ^2}{192 \epsilon ^2}
 -\tfrac{5 \zeta _3}{4 \epsilon }-\tfrac{199 \pi ^4}{11520}+\left(\tfrac{17 \pi ^2 \zeta _3}{24}
 -\tfrac{59 \zeta _5}{8}\right) \epsilon \right. \nn\\& \left. 
 +\left(\tfrac{15 \zeta _3^2}{2}-\tfrac{775 \pi ^6}{290304}\right) \epsilon ^2
 +\mathcal{O}\left(\epsilon ^3\right)\right), \\
 \widetilde{C}_{22,-3,0}(\epsilon ) &=  e^{i \pi  \epsilon } \left(\tfrac{11}{16 \epsilon ^4}
 -\tfrac{13 \pi ^2}{48 \epsilon ^2}-\tfrac{17 \zeta _3}{2 \epsilon }
 -\tfrac{83 \pi ^4}{480}+\left(\tfrac{35 \pi ^2 \zeta _3}{12}-\tfrac{387 \zeta _5}{8}\right) \epsilon 
 \right. \nn\\& \left.
 +\left(\tfrac{181 \zeta _3^2}{4}-\tfrac{3457 \pi ^6}{60480}\right) \epsilon ^2+\mathcal{O}\left(\epsilon ^3\right)\right), \\
 \widetilde{C}_{41,-3,0}(\epsilon ) &=  \tfrac{e^{i \pi  \epsilon }}{1-2 \epsilon }\left(-\tfrac{1}{8 \epsilon ^3}
 +\tfrac{\pi ^2}{12 \epsilon }+2 \zeta _3+\tfrac{53 \pi ^4 \epsilon }{1440}+\left(\tfrac{129 \zeta _5}{8}
 -\tfrac{29 \pi ^2 \zeta _3}{24}\right) \epsilon ^2  \right. \nn\\& \left. 
 +\left(\tfrac{341 \pi ^6}{20160}
 -\tfrac{55 \zeta _3^2}{4}\right) \epsilon ^3+\mathcal{O}\left(\epsilon ^4\right)\right), \\
 \widetilde{C}_{44,-2,0}(\epsilon ) &=  \tfrac{1}{1+\epsilon}\left(-\tfrac{1}{8 \epsilon ^3}
 +\tfrac{5 \pi ^2}{48 \epsilon }+\tfrac{9 \zeta _3}{4}+\tfrac{19 \pi ^4 \epsilon }{576}
 +\left(\tfrac{63 \zeta _5}{4}-\tfrac{11 \pi ^2 \zeta _3}{8}\right) \epsilon ^2  \right. \nn\\& \left.
 +\left(\tfrac{101 \pi ^6}{8064}-\tfrac{57 \zeta _3^2}{4}\right) \epsilon ^3+\mathcal{O}\left(\epsilon ^4\right)\right), \\
 \widetilde{C}_{45,-3,0}(\epsilon ) &=  e^{i \pi  \epsilon } \left(-\tfrac{3}{8 \epsilon ^4}
 +\tfrac{11 \pi ^2}{48 \epsilon ^2}+\tfrac{5 \zeta _3}{\epsilon }+\tfrac{199 \pi ^4}{2880}
 +\left(\tfrac{59 \zeta _5}{2}-\tfrac{17 \pi ^2 \zeta _3}{6}\right) \epsilon   \right. \nn\\& \left.
 +\left(\tfrac{775 \pi ^6}{72576}-30 \zeta _3^2\right) \epsilon ^2+\mathcal{O}\left(\epsilon ^3\right)\right), \\
 \widetilde{C}_{46,-3,0}(\epsilon ) &=  \tfrac{e^{i \pi  \epsilon }}{1+4 \epsilon}\left(-\tfrac{19}{32 \epsilon ^3}
 +\tfrac{19 \pi ^2}{64 \epsilon }+\tfrac{77 \zeta _3}{16}
 -\tfrac{49 \pi ^4 \epsilon }{1280}+\left(-\tfrac{77}{32} \pi ^2 \zeta _3-\tfrac{441 \zeta _5}{16}\right) \epsilon ^2
  \right. \nn\\& \left.
 +\left(-\tfrac{403 \zeta _3^2}{16}-\tfrac{21953 \pi ^6}{96768}\right) \epsilon ^3+\mathcal{O}\left(\epsilon ^4\right)\right), \\
 \widetilde{C}_{47,-2,0}(\epsilon ) &=  \tfrac{1}{1-2 \epsilon }\left(-\tfrac{3}{8 \epsilon ^3}+\tfrac{\pi ^2}{6 \epsilon }
 +\tfrac{7 \zeta _3}{2}+\tfrac{\pi ^4 \epsilon }{20}+\left(16 \zeta _5-\tfrac{4 \pi ^2 \zeta _3}{3}\right) \epsilon ^2
 \right. \nn\\& \left.
 +\left(\tfrac{191 \pi ^6}{22680}-13 \zeta _3^2\right) \epsilon ^3+\mathcal{O}\left(\epsilon ^4\right)\right), \\
 \widetilde{C}_{49,-3,0}(\epsilon ) &=  
 e^{i \pi  \epsilon } \Big(
 -\tfrac{1}{2 \epsilon ^4}
 +\tfrac{1}{2 \epsilon ^3}
 +\tfrac{1}{\epsilon ^2}\left( -\tfrac{3}{2}-\tfrac{\pi ^2}{4} \right)
 +\tfrac{1}{\epsilon }\left( -6 \zeta _3-\pi ^2+\tfrac{9}{2}\right)
 \nn\\&
 +\left(-23 \zeta _3-\tfrac{43 \pi ^4}{240}+3 \pi ^2-\tfrac{27}{2}\right)
 +\left(\tfrac{25 \pi ^2 \zeta _3}{2}+69 \zeta _3-\tfrac{329 \zeta _5}{2}-\tfrac{19 \pi ^4}{40}-9 \pi ^2+\tfrac{81}{2}\right) \epsilon 
 \nn\\&
 +\left(142 \zeta _3^2+19 \pi ^2 \zeta _3-207 \zeta _3-252 \zeta _5-\tfrac{51 \pi ^6}{224}+\tfrac{57 \pi ^4}{40}+27 \pi ^2-\tfrac{243}{2}\right) \epsilon ^2
\nn\\&
 +\mathcal{O}\left(\epsilon ^3\right)\!
 \Big),\\ 
 \widetilde{C}_{60,-3,0}(\epsilon ) &=  e^{i \pi  \epsilon } \left(\tfrac{1}{16 \epsilon ^4}
 -\tfrac{\pi ^2}{32 \epsilon ^2}-\tfrac{7 \zeta _3}{8 \epsilon }-\tfrac{31 \pi ^4}{1920}
 +\left(\tfrac{7 \pi ^2 \zeta _3}{16}-\tfrac{45 \zeta _5}{8}\right) \epsilon \right. \nn\\& \left.
 +\left(\tfrac{49 \zeta _3^2}{8}-\tfrac{53 \pi ^6}{11520}\right) \epsilon ^2
 +\mathcal{O}\left(\epsilon ^3\right)\right), \\
 \widetilde{C}_{67,-2,0}(\epsilon ) &=  \left(\tfrac{1}{4 \epsilon ^4}-\tfrac{5 \pi ^2}{24 \epsilon ^2}
 -\tfrac{9 \zeta _3}{2 \epsilon }
 -\tfrac{19 \pi ^4}{288}+\left(\tfrac{11 \pi ^2 \zeta _3}{4}-\tfrac{63 \zeta _5}{2}\right) \epsilon \right. \nn\\&\left.
 +\left(\tfrac{57 \zeta _3^2}{2}-\tfrac{101 \pi ^6}{4032}\right) \epsilon ^2+\mathcal{O}\left(\epsilon ^3\right)\right),\\
 \widetilde{C}_{86,-3,0}(\epsilon ) &=  e^{i \pi  \epsilon } \left(\tfrac{3}{8 \epsilon ^4}
 -\tfrac{11 \pi ^2}{48 \epsilon ^2}-\tfrac{5 \zeta _3}{\epsilon }
 -\tfrac{199 \pi ^4}{2880}+\left(\tfrac{17 \pi ^2 \zeta _3}{6}-\tfrac{59 \zeta _5}{2}\right) \epsilon 
 \right. \nn\\& \left.
 +\left(30 \zeta _3^2-\tfrac{775 \pi ^6}{72576}\right) \epsilon ^2+\mathcal{O}\left(\epsilon ^3\right)\right), \\
 \widetilde{C}_{87,-3,0}(\epsilon ) &=  e^{i \pi  \epsilon } \left(-\tfrac{5}{16 \epsilon ^4}
 +\tfrac{11 \pi ^2}{32 \epsilon ^2}+\tfrac{11 \zeta _3}{\epsilon }
 +\tfrac{193 \pi ^4}{640}+\left(\tfrac{549 \zeta _5}{4}
 -\tfrac{17 \pi ^2 \zeta _3}{2}\right) \epsilon \right. \nn\\& \left.
 +\left(\tfrac{47227 \pi ^6}{241920}-118 \zeta _3^2\right) \epsilon ^2+\mathcal{O}\left(\epsilon ^3\right)\right), \\
 \widetilde{C}_{121,-3,0}(\epsilon ) &=  e^{i \pi  \epsilon } \left(\tfrac{5}{16 \epsilon ^4}
 -\tfrac{11 \pi ^2}{32 \epsilon ^2}-\tfrac{11 \zeta _3}{\epsilon }-\tfrac{193 \pi ^4}{640}
 +\left(\tfrac{17 \pi ^2 \zeta _3}{2}-\tfrac{549 \zeta _5}{4}\right) \epsilon \right. \nn\\& \left.
 +\left(118 \zeta _3^2-\tfrac{47227 \pi ^6}{241920}\right) \epsilon ^2+\mathcal{O}\left(\epsilon ^3\right)\right), \\
 \widetilde{C}_{122,-3,0}(\epsilon ) &=  e^{i \pi  \epsilon } \left(-\tfrac{11}{32 \epsilon ^4}
 +\tfrac{13 \pi ^2}{96 \epsilon ^2}+\tfrac{17 \zeta _3}{4 \epsilon }
 +\tfrac{83 \pi ^4}{960}+\left(\tfrac{387 \zeta _5}{16}-\tfrac{35 \pi ^2 \zeta _3}{24}\right) \epsilon 
 \right. \nn\\& \left.
 +\left(\tfrac{3457 \pi ^6}{120960}-\tfrac{181 \zeta _3^2}{8}\right) \epsilon ^2+\mathcal{O}\left(\epsilon ^3\right)\right), \\
 \widetilde{C}_{126,-3,0}(\epsilon ) &=  \tfrac{e^{i \pi  \epsilon }}{1+4 \epsilon}\left(-\tfrac{19}{32 \epsilon ^3}
 +\tfrac{19 \pi ^2}{64 \epsilon }+\tfrac{77 \zeta _3}{16}-\tfrac{49 \pi ^4 \epsilon }{1280}
 +\left(-\tfrac{77}{32} \pi ^2 \zeta _3-\tfrac{441 \zeta _5}{16}\right) \epsilon ^2 \right. \nn\\& \left.
 +\left(-\tfrac{403 \zeta _3^2}{16}-\tfrac{21953 \pi ^6}{96768}\right) \epsilon ^3+\mathcal{O}\left(\epsilon ^4\right)\right).
\end{align}

In the following subsections we provide some examples of the calculation of some of the constants above.
All other constants can be obtained by suitable extensions of the computations presented below.

\subsection{Boundary integral $B_1$}

Boundary integral $B_1$ is one of the simplest integrals and can be computed exactly in $\eps$.
Its integral representation is given by 
\begin{align}
B_1 =
\int 
\frac{d^d k_1}{(2\pi)^{d-1}}
\frac{d^d k_2}{(2\pi)^{d-1}}
\frac{d^d k_3}{(2\pi)^{d}}
\frac{
\delta^{+\!\!}\left(k_1^2\right)
\delta^{+\!\!}\left(k_2^2\right)
\delta\left(p \cdot k_{12} - \tfrac{1}{2}\right)
\delta\left(\pbar \cdot k_{12} - \tfrac{1-z}{2} \right)
}{
k_{13}^2 \, (p-k_{123})^2
}\,.
\label{eq:B1v1}
\end{align}
The integral over $k_3$ is  performed first. 
In this case it is a simple one-loop bubble integral given in \cref{eq:Bub1}.
For convenience, we also introduce the abbreviation
\begin{align}
\mathfrak{D}k_{12}
&=
\frac{d^d k_1}{(2\pi)^{d-1}}
\frac{d^d k_2}{(2\pi)^{d-1}}
\delta^{+\!\!}\left(k_1^2\right)
\delta^{+\!\!}\left(k_2^2\right)
\delta\left(p \cdot k_{12} - \tfrac{1}{2}\right)
\delta\left(\pbar \cdot k_{12} - \tfrac{1-z}{2}\right).
\label{eq:k12measure}
\end{align}
After these steps the boundary integral is written as
\begin{align}
B_1 
= 
C_{\text{bub}}(\eps)
\int \mathfrak{D}k_{12}
\left( - (p-k_2)^2 \right)^{-\eps}
= 
C_{\text{bub}}(\eps)
\int \mathfrak{D}k_{12}
\left( 2 p \cdot k_2 \right)^{-\eps}
\,,
\label{eq:B1v2}
\end{align}
where we have used the on-shell conditions $p^2 = k_2^2 = 0$. 
The prefactor $C_{\text{bub}}(\eps)$ is defined below \cref{eq:Bub1}.

We proceed by writing the remaining integration measure in the form
\begin{align}
\int \mathfrak{D}k_{12}
&=
\frac{1}{4 (2\pi)^{2d-2}}
\left(
\prod_{i=1}^{2}
\int d\alpha_i \, d\beta_i \, d\Omega^{(i)}_{d-2}~
(\alpha_i \beta_i)^{-\eps}
\right)
\delta\left(\beta_{12} - 1\right)
\delta\left(\alpha_{12} - (1-z)\right),
\label{eq:SudaMeasure}
\end{align}
which is convenient for extracting the leading behaviour in the limit $z \to 1$.
The expression in \cref{eq:SudaMeasure} is obtained by inserting the Sudakov decomposition \cref{eq:Sudakov} into  \cref{eq:k12measure} and integrating out the length of the vector $k_{i \perp}$. 
Accordingly, one should set $2 k_i \cdot \pbar = \alpha_i$, $2 k_i \cdot p = \beta_i$ and $k_{12}^2 = 2 k_1 \cdot k_2 = (\alpha_1 \beta_2 + \alpha_2 \beta_1 - 2\sqrt{\alpha_1 \alpha_2 \beta_1 \beta_2} \cos \theta_{12})$ in the integrand when using the measure in \cref{eq:SudaMeasure}. 

Upon inserting this integration measure into \cref{eq:B1v2} we obtain
\begin{align}
B_1 
&= 
\frac{C_{\text{bub}}(\eps)}{4 (2\pi)^{2d-2}}
\left(
\prod_{i=1}^{2}
\int d\alpha_i \, d\beta_i \, d\Omega^{(i)}_{d-2}
\right)
\alpha_1^{-\eps}
\alpha_2^{-\eps}
\beta_1^{-\eps}
\beta_2^{-2\eps}
\delta\left(\beta_{12} - 1\right)
\delta\left(\alpha_{12} - (1-z)\right).
\label{eq:B1v3}
\end{align}
Since the integrand does not depend on $k_{12}^2$, the angular integrations $d\Omega_{d-2}^{(1)}$ and $d\Omega_{d-2}^{(2)}$ are trivial.
The integrations over $\alpha_{1}, \alpha_{2}, \beta_{1}, \beta_{2}$ are performed using
\begin{align}
\int d\xi_1 d\xi_2 ~ \xi_1^{a}\, \xi_2^{b} \, \delta\left(\xi_{12}-X\right) = \frac{\Gam(1+a)\Gam(1+b)}{\Gam(2+a+b)}\,X^{1+a+b} \,.
\label{eq:paramint}
\end{align}

As a result, we obtain
\begin{align}
B_1 = 
\left(\frac{\Omega_{d-2}}{(2\pi)^{d-1}}\right)^3 
\frac{i}{16} 
\frac{\Gam(\eps)\Gam(1-\eps)^6\Gam(1-2\eps)}{\Gam(2-2\eps)^2\Gam(2-3\eps)} 
(1-z)^{1-2\eps}\,.
\end{align}
Here, $\Omega_{d-2} = 2 \pi^{1-\eps} / \Gam(1-\eps)$.
The prefactor of $(1-z)^{1-2\eps}$ is the desired constant $C_{1,-2,0}(\eps)$.

\subsection{Boundary integral $B_2$}

Our second example concerns the   boundary integral $B_2$. It reads
\begin{align}
B_2
= \int \mathfrak{D}k_{12}  \int \frac{d^d k_3}{(2\pi)^{d}}~\frac{1}{k_{3}^2 \, k_{123}^2}
= C_{\text{bub}}(\eps) \int \mathfrak{D}k_{12}\,( - k_{12}^2 )^{-\eps} \,.
\label{eq:B2v1}
\end{align}
In this case the integrand depends on $k_{12}^2$, and therefore the Sudakov decomposition of the momenta $k_1$ and $k_2$ leads to non-trivial angular integral $d\Omega^{(i)}_{d-2}$ in \cref{eq:SudaMeasure}.
With this example we demonstrate how that problem can be treated, at least in cases when the loop integral gives a relatively simple result.

We start by inserting the identity $1 = \int d^dQ \,\delta^{d}(k_{12}-Q)$, which has the effect of factorizing the $k_{12}^2$-dependence.
The boundary integral is then written as
\begin{align}
B_2
= 4 C_{\text{bub}}(\eps) \int d^dQ \,\delta(2 p \cdot Q - 1)\,\delta(2 \pbar \cdot Q - (1-z))\,( - Q^2)^{-\eps} \, \mathrm{PS}(Q)\,,
\label{eq:B2v2}
\end{align}
where $\mathrm{PS}(Q)$ is the standard two-particle massless phase-space integral \cite{Gehrmann-DeRidder:2003pne}
\begin{align}
\mathrm{PS}(Q) 
= \int d^d k_1 d^d k_2 \, \delta^{+\!}(k_1^2) \,\delta^{+\!}(k_2^2)\, \delta^{d}(k_{12}-Q)
= \frac{\Omega_{d-2}}{4} \frac{\Gam(1-\eps)^2}{\Gam(2-2\eps)} (Q^2)^{-\eps}\,.
\end{align}

Next, we proceed by making a Sudakov decomposition of $Q$ analogous to \cref{eq:Sudakov}. 
That produces a parametric integral of the form
\begin{align}
\mathcal{I} 
&= \int d\alpha\, d\beta\, d(Q_\perp^2)\, (Q_{\perp}^2)^{-\eps} \,\delta(\beta - 1)\,\delta(\alpha - (1-z))\,( \alpha \beta - Q_{\perp}^2 )^{-2\eps} \\
&=\int_0^{1-z} d(Q_\perp^2)\, (Q_{\perp}^2)^{-\eps} \,( (1-z) - Q_{\perp}^2 )^{-2\eps} 
= \frac{\Gam(1-\eps)\Gam(1-2\eps)}{\Gam(2-3\eps)} (1-z)^{1-3\eps}\,,
\end{align}
where the bounds on the integral over $Q_\perp^2$ are dictated by the conditions $Q^2 = \alpha \beta - Q_{\perp}^2 > 0$ and $Q_{\perp}^2>0$.
As a result, we find
\begin{align}
B_2 = 
\left(\frac{\Omega_{d-2}}{(2\pi)^{d-1}}\right)^3 
\frac{i e^{i \pi \epsilon}}{16} 
\frac{\Gam(\eps)\Gam(1-\eps)^6\Gam(1-2\eps)}{\Gam(2-2\eps)^2\Gam(2-3\eps)} 
(1-z)^{1-3\eps}\,.
\label{eq:B2v3}
\end{align}

\vspace{5mm}
Other boundary integrals, that can be computed via the same method, lead to a two-particle massless phase-space integral of the form
\begin{align}
\mathrm{PS}_n(Q, p, \pbar) 
= \int d^d k_1 d^d k_2 \, \frac{\delta^{+\!}(k_1^2) \,\delta^{+\!}(k_2^2)\, \delta^{d}(k_{12}-Q)}{(2 k_1 \cdot \pbar) (2 k_2 \cdot p)^n}~~\text{for}~n \geq 0,~ p^2 = \pbar^2 = 0\,.
\label{eq:PSN}
\end{align}
One way to calculate the phase-space integral in \cref{eq:PSN} is to work in the rest frame $Q = (Q_0,\vec{0})$, to carry out the resulting angular integrations (see eq.~(49) in ref.~\cite{Somogyi:2011ir}), and to write the expression back in a Lorentz invariant form.
The result is
\begin{align}
\mathrm{PS}_n(Q, p, \pbar) &= 
\frac{\Omega_{d-2}}{4} 
\frac{\Gam(-\eps) \Gam(1-n-\eps)}{\Gam(1-n-2\eps)} 
\frac{(Q^2)^{-\eps}}{(2 \pbar \cdot Q) \, (2 p \cdot Q)^n} ~
\nn\\&\qquad
{}_2F_1\left(1,n;1-\eps;\frac{Q^2 (2 p\cdot \pbar)}{(2 p \cdot Q) (2 \pbar \cdot Q)}\right).
\label{eq:PSNresult}
\end{align}

\subsection{Boundary integral $B_8$}

As the next example, we consider the boundary integral $B_8$. It is given by 
\begin{align}
B_8 = \int \frac{\mathfrak{D}k_{12}}{(k_1 \cdot \pbar)}  
\int \frac{d^d k_3}{(2\pi)^{d}}~\frac{1}{k_{3}^2 \, k_{13}^2 \, k_{123}^2 \, (p-k_{123})^2}\,. 
\label{eq:B8-v1}
\end{align}

The result of the one-loop one-mass box integral is given as $\text{Box1}$ in \cref{eq:Box1-v2}.
In order to find the behaviour of Box1 in the limit $z \to 1$, it is convenient to rewrite the last two hypergeometric functions in \cref{eq:Box1-v2}
using  the identity
\begin{align}
{}_2F_1(a,b;c;z) 
= \frac{\Gam(b-a)\Gam(c)}{\Gam(b)\Gam(c-a)} (-z)^{-a} 
{}_2F_1(a,a-c+1;a-b+1;1/z) + \{ a \leftrightarrow b \}\,.
\end{align}
As a result, the last two terms in \cref{eq:Box1-v2} combine to produce a contribution 
that is sub-leading with respect to the first term in the limit $z \to 1$.
We are left with the calculation of the following integral 
\begin{align}
B_8|_{z \to 1}  = 
C_{\text{Box}}(\eps)
\int \frac{\mathfrak{D}k_{12}}{(k_1 \cdot \pbar) \, (2 k_2 \cdot p)} 
\left( -k_{12}^2 \right)^{-1-\eps} {}_2F_1\left(1,-\eps;1-\eps;-\frac{2 k_1 \cdot p}{2 k_2 \cdot p}\right) .
\label{eq:B8-v2}
\end{align}
Here the integrand depends on $k_{12}^2$ but, 
 unlike in the previous example, inserting $1 = \int d^dQ \,\delta^{d}(k_{12}-Q)$ will not be helpful, because that would lead to a two-particle phase-space integral whose integrand contains the hypergeometric function in \cref{eq:B8-v2}.
In such a situation there is no choice but to perform a non-trivial angular integration directly. 
In this example we demonstrate how to carry out such an integral.

We proceed by making a Sudakov decomposition of $k_1$ and $k_2$ in \cref{eq:B8-v2}.
The factor $(k_{12}^2)^{-1-\eps}$ in the integrand then leads to the following angular integral 
\begin{align}
&\int \frac{d\Omega_{d-2}^{(1)} }{(\alpha_1 \beta_2 + \alpha_2 \beta_1 - 2 \sqrt{\alpha_1 \alpha_2 \beta_1 \beta_2 } \cos\theta_{12})^{1+\eps}} 
\nn\\[2mm]
&= 
\Omega_{d-3} \int_{0}^{\pi} \frac{d\theta_{12} (1-\cos^2\theta_{12})^{-\eps}}{(\alpha_1 \beta_2 + \alpha_2 \beta_1 - 2 \sqrt{\alpha_1 \alpha_2 \beta_1 \beta_2} \cos\theta_{12})^{1+\eps}} 
\nn\\[1mm]
&=
\Omega_{d-3} \,
4^{-\eps} \,
\frac{\Gam(\tfrac{1}{2}-\eps)^2}{\Gam(1-2\eps)} \,
\frac{{}_2F_1\left(1+\eps,\tfrac{1}{2}-\eps;1-2\eps;\frac{4 \sqrt{\alpha_1 \alpha_2 \beta_1 \beta_2}}{\left(\sqrt{\alpha_1\beta_2}+\sqrt{\alpha_2\beta_1}\right)^2}\right)}{\left(\sqrt{\alpha_1\beta_2}+\sqrt{\alpha_2\beta_1}\right)^{2+2\eps}}\,.
\label{eq:angularintegrals}
\end{align}

Subsequently, we rescale $\alpha_i \to (1-z) \alpha_i$. 
This produces the overall scaling of the integral $(1-z)^{-1-3\eps}$ and changes the constraints on the $\alpha$'s into $\delta(\alpha_{12}-1)$.
Integrating over the delta functions, we obtain $\alpha_2=1-\alpha_1$ and $\beta_2=1-\beta_1$.
The remaining two-fold integration over $\alpha_1$ and $\beta_1$ must then be split into two pieces: (i) $\alpha_1 > \beta_1$, and (ii) $\alpha_1 < \beta_1$, in order to simplify the argument of the hypergeometric function.
In case (i) we have that $\alpha_1 \beta_2 > \alpha_2 \beta_1$, which allows us to rewrite
\begin{align}
\frac{{}_2F_1\left(1+\eps,\tfrac{1}{2}-\eps;1-2\eps;\frac{4 \sqrt{\alpha_1 \alpha_2 \beta_1 \beta_2}}{\left(\sqrt{\alpha_1\beta_2}+\sqrt{\alpha_2\beta_1}\right)^2}\right)}{\left(\sqrt{\alpha_1\beta_2}+\sqrt{\alpha_2\beta_1}\right)^{2+2\eps}}
&=
\frac{{}_2F_1\left(1,\tfrac{1}{2}
-\eps;1-2\eps;\frac{4 \xi}{\left(1+\xi\right)^2}\right)}{(\alpha_1 \beta_2)^{1+\eps} \left(1+\xi\right)^{2+2\eps}} 
\nn\\
&=
\frac{{}_2F_1\left(1+\eps,1+2\eps;1-\eps;\xi^2\right)}{(\alpha_1 \beta_2)^{1+\eps}}\,,
\end{align}
where $\xi=\sqrt{\frac{\alpha_2 \beta_1}{\alpha_1 \beta_2}}$ and $|\xi|<1$.
Case (ii) is completely analogous, but with $\xi=\sqrt{\frac{\alpha_1 \beta_2}{\alpha_2 \beta_1}}$.

Following the above discussion, we write the integral $B_8$ as the sum of two terms 
\begin{align}
B_8 = 
- C_{\text{Box}}(\eps)
\frac{\Omega_{d-2}\Omega_{d-3}}{(2\pi)^{2d-2}}
\frac{e^{i \pi \eps}}{2^{1+2\eps}}
\frac{\Gam(\tfrac{1}{2}-\eps)^2}{\Gam(1-2\eps)}
(1-z)^{-1-3\eps}
\left( X_{8}^{\text{(i)}} + X_{8}^{\text{(ii)}} \right).
\end{align}
The contribution from case (i) to the integral is
\begin{align}
X_{8}^{\text{(i)}} 
&= 
\int_0^1 d\alpha_1 \, d\beta_1 \, \theta(\alpha_1-\beta_1)
(\alpha_2 \beta_1 )^{-\eps}\,
(\alpha_1 \beta_2 )^{-2-2\eps}\,
\nn\\&\hspace{5mm}\times 
{}_2F_1\left(1,-\eps;1-\eps;-\frac{\beta_1}{\beta_2}\right)\,
{}_2F_1\left(1+\eps,1+2\eps;1-\eps;\frac{\alpha_2 \beta_1}{\alpha_1 \beta_2}\right).
\end{align}
where $\alpha_2 = 1-\alpha_1$ and $\beta_2 = 1-\beta_1$.
After a change of variables $\beta_1 \to r$ with $r = \frac{\alpha_2 \beta_1}{\alpha_1 \beta_2}$ and, subsequently,  $\alpha_1 \to t $ with $\alpha_1 = \frac{t}{r+t-r t}$, it becomes 
\begin{align}
X_{8}^{\text{(i)}} 
&= \int_0^1 dr \,dt \,\,
r^{-\eps}\,
t^{-1-3\eps}\,
(1-t)^{-1-3\eps}\,
(r+t-rt)^{3\eps}\,
\nn\\&\hspace{5mm}\times 
{}_2F_1\left(1+\eps,1+2\eps;1-\eps;r\right)
{}_2F_1\left(1,-\eps;1-\eps;\frac{t}{t-1}\right).
\end{align}
After applying identities for hypergeometric functions to simplify their argument and extract their singularities at the endpoints of
the integration, we obtain
\begin{align}
X_{8}^{\text{(i)}} 
&= \int_0^1 dr \,dt \,\,
r^{-\eps}\,
(1-r)^{-1-4\eps}
t^{-1-3\eps}\,
(1-t)^{-1-4\eps}\,
(r+t-rt)^{3\eps}\,
\nn\\&\hspace{5mm}\times 
{}_2F_1\left(-3\eps,-2\eps;1-\eps;r\right)
{}_2F_1\left(-\eps,-\eps;1-\eps;t\right).
\end{align}
The integrand has singularities at points $r=1$ and $t=0,1$.
The integral may be carried out by performing suitable subtractions at this points that enables expansion of complicated
integrals in $\epsilon$.   This procedure is tedious but relatively standard and its explanation is thus omitted here.
The calculation of $X_8^{\text{(ii)}}$ can be performed along the same lines. 
The final result for this boundary integral reads 
\begin{align}
B_8 &= 
\left(\frac{\Omega_{d-2}}{(2\pi)^{d-1}}\right)^3 
i e^{i \pi \epsilon} 
(1-z)^{-1-3\eps} 
\Bigg[
-\frac{5}{16 \eps^4}
+\frac{11 \pi^2}{32 \eps^2}
+\frac{11 \zeta_3}{\eps}
+\frac{193 \pi^4}{640}
\nn\\&\hspace{20mm}
+\left(\frac{549 \zeta_5}{4}-\frac{17 \pi ^2 \zeta_3}{2}\right) \eps 
+\left(\frac{47227 \pi ^6}{241920}-118 \zeta_3^2\right) \eps^2
+\order{\eps^3}
\Bigg].
\end{align}


\section{Numerical checks of master integrals}
\label{sec:checks}
The calculation of the master integrals required many non-trivial steps and therefore it 
is good to have a completely independent check of our results for the integrals. There are various public codes that can evaluate 
loop integrals numerically, but none as of yet that can compute 
phase-space integrals,  especially of the type that we consider 
in this paper.  The complication arises from integration over angles 
of the emitted gluons since it is challenging to find a suitable 
parametrization for the angular degrees of freedom. 

One possibility, pointed 
out in ref.~\cite{Anastasiou:2013srw}, is to use 
the  Mellin-Barnes (MB) representation for this purpose. The idea is to 
split complex denominators that appear in integrals into integrals 
of products of simpler scalar products, perform ensuing 
angular integrals analytically using 
results of ref.~\cite{Somogyi:2011ir} and   compute the resulting 
MB integrals numerically using available MB packages~\cite{MBTools}. 

Taking, as an example, a propagator 
$((p-k_{12})^2)^{-1}=2^{-1}(k_1\cdot k_2-p\cdot k_1-p\cdot k_2)^{-1}$,  
we split into an integral of products of $k_1 \cdot k_2$, 
$p \cdot k_1$ and $p \cdot k_2$ by repeatedly applying 
the MB representation
\begin{align}
\frac{1}{(x+y)^\lambda}=\int \limits_{-i\infty}^{+i\infty} \frac{dz}{2\pi i} \frac{y^z}{x^{z+\lambda}}\frac{\Gamma(-z)\Gamma(\lambda+z)}{\Gamma(\lambda)}\,.
\label{eq:MB}
\end{align}
In \cref{eq:MB} the contour has to be  chosen in such a way 
that the poles of $\Gamma(-z)$ are to 
the right and the poles of $\Gamma(\lambda+z)$ are to the left 
of the contour that runs along the imaginary axis. 
However,  if either $x$ or $y$ is negative in \cref{eq:MB}, 
the numerical evaluation of the right hand side of \cref{eq:MB} 
may become unstable because of the exponential 
increase of $(-x-i0)^z$ as $\text{Im}[z]\rightarrow\infty$. 
This implies that if we would split the denominator 
$(k_1\cdot k_2-p\cdot k_1-p\cdot k_2)^{-1}$ into MB integrals,  numerical integration may become
unstable.\footnote{We remind the reader that $k_1\cdot k_2, p\cdot k_1, p\cdot k_2\geq 0$.}

It is possible to get around this problem by considering a {\it decay} process 
instead of the production process. Indeed, 
our phase-space integrals correspond to an 
incoming parton emitting two collinear particles 
before entering a hard process; since the incoming parton has zero invariant 
mass, the off-shellness of a quark line becomes negative after gluon emissions. 
If, on the other hand, a quark with positive virtuality {\it leaves} the hard 
process and decays to a zero-virtuality final state quark by emitting 
gluons, virtual quark lines at intermediate stages have positive virtualities 
for which numerical integration of the relevant 
 MB representations is straightforward. 
In order to get from a production kinematics to a decay kinematics, 
we need to change the four-momenta 
$p\rightarrow -p, \bar{p}\rightarrow -\bar{p}$. The constraint 
on the longitudinal momentum of a virtual quark in the decay kinematics 
becomes  $k_{12}\cdot \bar{p} = (z-1)$. Since this quantity should be positive, 
we have to take $z \geq  1$.  
The virtuality constraint reads  
$k_{12}\cdot p = -t/(2z)$. Since $k_{12} \cdot p$ is positive definite,  
we have to take $t\leq 0$. 

For the sake of example, consider an integral in the decay kinematics
\begin{align}
I_{\text{decay}}(\kappa,z) &=& \int\frac{d^d k_1}{(2\pi)^{d-1}}\frac{d^d k_2}{(2\pi)^{d-1}}\frac{d^d k_3}{(2\pi)^{d}}\frac{\delta^{+}\!\left(k_1^2\right)\delta^{+}\!\left(k_2^2\right)\delta\left(k_{12}\cdot p+\frac{\kappa}{2}\right)\delta\left(k_{12}\cdot \bar{p}+\frac{(1-z)}{2}\right)}{(k_3^2)^{a_1}((p+k_1)^2)^{a_2}((p+k_{123})^2)^{a_3}(\bar{p}\cdot k_1)^{a_4}}\,, 
\label{eq:Ibargen}
\end{align}
where $\kappa=t/z$ and $2p\cdot\bar{p}=1$. Its analytic expression 
can be found from our 
solutions for integrals in the production channel
\begin{align}
I_{\text{production}}(\kappa,z) = \int\frac{d^d k_1}{(2\pi)^{d-1}}\frac{d^d k_2}{(2\pi)^{d-1}}\frac{d^d k_3}{(2\pi)^{d}}\frac{\delta^{+}\!\left(k_1^2\right)\delta^{+}\!\left(k_2^2\right)\delta\left(k_{12}\cdot p-\frac{\kappa}{2}\right)\delta\left(k_{12}\cdot \bar{p}-\frac{(1-z)}{2}\right)}{(k_3^2)^{a_1}((p-k_1)^2)^{a_2}((p-k_{123})^2)^{a_3}(\bar{p}\cdot k_1)^{a_4}} \label{eq:Igen}
\end{align}
by   an analytic  continuation of $\kappa$ and $z$ to the region 
$\kappa\leq 0, z\geq 1$.\footnote{Additional multiplication by 
$(-1)^{a_4}$ is needed as well.}
Note that the variable dependence 
$s=2p\cdot \bar{p}$ is unchanged when moving from production to decay kinematics.

The propagators in the decay kinematics, \cref{eq:Ibargen},
are given by sums of positive-definite quantities and, for this reason, 
are more suitable for the MB integration.  It is therefore 
more convenient to numerically compute integrals in decay 
kinematics, \cref{eq:Ibargen}, and compare them with 
analytically-continued  integrals computed in the production 
channel.

To implement this in practice, we note that 
if specific 
combination $\delta(1-2k_{1\ldots n}\cdot (p+\bar{p}))\prod_{i=1}^n 
\delta^{+}(k_i^2)$ of delta functions appears in the integrand, 
it is known how to perform  
phase-space integrals with the MB method~\cite{Anastasiou:2013srw}. 
In our case,  different delta-functions appear in integrands 
but we may produce such a combination of delta 
functions by integrating the function $I_{\text{decay}}(\kappa,z)$ over the variables $\kappa$ and $z$, both from 
$-\infty$ to $+\infty$, with an extra delta function insertion that 
imposes a further constraint $\kappa=z-2$. We obtain 
\begin{align}
N & = \int_1^2 dz \, I_{\text{decay}}(z-2,z) 
\nn\\
& = \int_{-\infty}^{+\infty} d\kappa \int_{-\infty}^{+\infty} dz \, I_{\text{decay}}(\kappa,z)\, \delta(z-2-\kappa) \nn\\
& =  4\int\frac{d^d k_1}{(2\pi)^{d-1}}\frac{d^d k_2}{(2\pi)^{d-1}}\frac{d^d k_3}{(2\pi)^{d}}\frac{\delta^{+}\!\left(k_1^2\right)\delta^{+}\!\left(k_2^2\right)\delta(1-2k_{12}\cdot (p+\bar{p}))}{(k_3^2)^{a_1}((p+k_1)^2)^{a_2}((p+k_{123})^2)^{a_3}(\bar{p}\cdot k_1)^{a_4}}\,. 
\label{eq:check}
\end{align}

The second equality in \cref{eq:check} follow from the fact that 
the decay kinematics imposes that $I_{\text{decay}}(\kappa,z)$ is exactly zero outside of the 
region $\kappa\leq 0, z\geq 1$. As we already mentioned, the first line in \cref{eq:check} 
can be evaluated starting from 
the  analytic solution for $I_{\text{production}}(t,z)$ and analytically 
continuing  from $t > 0$ to  $t = z-2 < 0$ 
and from $ 0 < z < 1$ to $ 1 < z < 2$.  
The integral that appears in the 
 last line \cref{eq:check}  
is a double real-virtual phase space integral 
that can be evaluated with the MB method.
By comparing the two results, we obtain  an indirect numerical check of our 
analytic solutions. 

We note that, by working with the decay kinematics, all one-loop 
virtual corrections have an imaginary part that, however, 
always factors out as an overall factor.  This can be seen from 
explicit expressions for the one-loop integrals shown in 
\cref{eq:Bub1,eq:Tri1,eq:Box1-v2,eq:Box2-v2,eq:Tri2-v2,eq:Box3-v2,eq:Pen1-v2}
when these integrals 
are written for the decay kinematics. 
For numerical checks,  we renormalize this overall factor 
away from both from the analytic result 
and from the  MB numerical computation.  The resulting 
integral is then real-valued which provides 
a good control on the  analytic continuation of the integrals 
from the production to 
decay kinematics.

We outline the steps that we take to evaluate phase-space integrals of the form given in \cref{eq:check}, 
adapting the line of reasoning given in~\cite{Anastasiou:2013srw} to the case of double-real virtual integrals. 

\begin{itemize}
 \item We begin by computing  the  one-loop integrals
over $k_3$ and express them 
in terms of a product of propagators with loop momenta $k_1,k_2$. 
For this we use formulas given in 
\cref{eq:Bub1,eq:Tri1,eq:Box1-v2,eq:Box2-v2,eq:Tri2-v2,eq:Box3-v2,eq:Pen1-v2} 
for various types of one-loop  integrals over $k_3$ with the mapping 
$p\rightarrow -p$. There are three types of 
 bubble integrals which are proportional to 
 $(-q^2)^{-\epsilon}=e^{i\pi\epsilon}(q^2)^{-\epsilon}$ 
with $q=k_{12},(p+k_2),(p+k_{12})$. The one-loop  triangles 
 and boxes are expressed as a sum of several terms that are evaluated 
separately. The integrals with the linear propagator $k_3.\bar{p}$ may be expressed in terms of MB integrals, after introducing 
MB variables in such a way that the integrals over the Feynman representation variables can be performed. For 
  some of the virtual integrals we need to use 
the MB representation of the hypergeometric function
 \begin{align}
{}_2F_1\left(a,b;c;x\right)=\frac{\Gamma(c)}{\Gamma(a)\Gamma(b)}\int^{+i\infty}_{-i\infty}\frac{dz}{2\pi i}
\frac{\Gamma(a+z)\Gamma(b+z)\Gamma(-z)}{\Gamma(c+z)}(-x)^z\,, 
\label{eq:2F1MB}
\end{align} 
 whenever $x<0$. Since $k_1\cdot k_2, p\cdot k_i>0$ one can check that the argument of the hypergeometric functions 
 that appear are always negative and the 
 MB representation in \cref{eq:2F1MB} is valid. The contour is again chosen such that the singularities of $\Gamma(\ldots -z)$ ($\Gamma(\ldots +z)$) 
are to the right (left) of the integration contour that runs  
along the imaginary axis. 
As we already mentioned, we 
extract and remove  overall factors of 
$ie^{i\pi\epsilon}$ that arise from the evaluation of one-loop integrals. 
After this step we are left with a double-real phase space integral over 
 $k_1,k_2$ that is a real-valued number.

 \item We express all the rational functions of scalar products of gluon 
and reference momenta, 
that arise  from the previous step through 
 products of simple scalar products 
 $k_{12}^2=2k_1\cdot k_2>0$, $(p+k_i)^2=2p\cdot k_i>0$ and $(\bar{p}+k_i)^2=2\bar{p}\cdot k_i>0$ and integrals over MB paramaters, repeatedly using the 
MB representation  \cref{eq:MB}. 
After that, integrals are written in the following symbolic form 
 \begin{align}
N \sim \sum_{k_3} \int\{dz_l\}|_{MB}\, \prod_{k=1}^m s_{ij}^{-\alpha_k}\,, \label{eq:MBstep2}
\end{align}
 where we have left out the remaining  integrations over $k_1,k_2$ that still need to be performed. The $\int\{dz_l\}|_{MB}$ factor represents the MB integrations 
 that arise from the hypergeometric function and those that have been introduced in order to split up 
the propagators of $k_1,k_2$  into two-particle invariants. 
 The sum $\sum_{k_3}$ indicates 
 that upon integration over $k_3$ several terms may arise that  
have to be  treated separately.

\item The rest of the calculation proceeds in full analogy with 
ref.~\cite{Anastasiou:2013srw} and we refer to that paper for further  details. The phase-space integration over 
energies of $k_1$ and $k_2$ 
is straightforward and the integration over 
angles can be performed using results presented in 
ref.~\cite{Somogyi:2011ir}.  Finally, we obtain integrals 
over MB parameters that have to be evaluated numerically. We 
use the package MB-tools~\cite{MBTools} for the numerical integration. 

\end{itemize} 

We compared the analytic integration of $N$ in \cref{eq:check} using our 
analytically-continued results for $I_{\text{decay}}(\kappa,z)$, with the numerical 
evaluation of $N$ by the method of MB and we found agreement.


\section{Results}
\label{sec:results}

In this section we present the results of the calculation.  
First, we provide the results for the 128 masters integrals that are listed in \cref{tab:listofmasters} in an ancillary file, which is available on \url{https://www.ttp.kit.edu/_media/progdata/2018/ttp18-034.tar.gz}
The results are organized as follows.
We provide the expressions for the master integrals in terms of canonical master
integrals, schematically $\vec{\mathcal{I}}(z,\eps) = \hat T(z,\eps)\,\vec{\mathcal{I}}_{\text{can}}(z,\eps)$. 
We also give  the 128 canonical master integrals $\vec{\mathcal{I}}_{\text{can}}(z,\eps)$ 
as  linear combinations of $z$-independent constants $C_{i}(\eps)$ multiplied by Taylor expansions
in $\eps$ that contain multiple polylogarithms $G_{\vec{a}}(z)$ up to weight 6.
The constants $C_{i}(\eps)$ are provided separately as Laurent expansions in $\eps$.
\vspace{5mm}

Our results are ingredients to the computation of the double-real contribution
to third-order matching coefficients $I_{ij}(t,z,\mu)$ of quark and gluon beam functions 
in perturbative QCD.
To illustrate this, we focus on the quark-to-quark branching process, i.e. $i=q$ and $j=q$, and write the perturbative expansion as
\begin{align}
I_{qq}(t,z,\mu) = \sum_{n=0}^{\infty} \left(\frac{\alpha_{s}}{4 \pi}\right)^n \,\, I^{(n)}_{qq}(t,z,\mu)\,.
\label{eq:matchcoefpertexp}
\end{align}
The N${}^3$LO term $I^{(3)}_{qq}(t,z,\mu)$ receives three contributions: single-real,  double-real and triple-real.
The master integrals calculated in this paper can be used to compute the double-gluon emission contribution to the matching coefficient. 
Restoring the normalization, we write  
\begin{align}
I^{(3),\text{RRV}}_{qq}(t,z,\mu) 
= \frac{1}{\mu^2} \left( \frac{t}{\mu^2} \right)^{-1-3\eps} 
\left(\frac{e^{\eps\gamma_E}}{\Gamma(1-\eps)}\right)^3  
I^{(3),\text{RRV}}_{qq}(z,\eps)\,.
\label{eq:matchcoeffact}
\end{align}
Its dependence on $t, z$ and $\mu$ factorises as expected.
The $t$-dependent factor is expanded in terms of plus distributions according to the formula
\begin{align}
x^{-1+k\eps}
= \frac{1}{k \eps}\delta\left(x\right) + \sum_{n \geq 0} \frac{(k\eps)^n}{n!} \mathcal{L}_n\left(x\right)\,.
\label{eq:tdistrib}
\end{align}

The non-trivial $z$-dependent factor on the right-hand side of \cref{eq:matchcoeffact} may be split into a part that diverges in the soft limit $z \to 1$ and a finite remainder
\begin{align}
I^{(3),\text{RRV}}_{qq}(z,\eps) = I^{(3),\text{RRV,div}}_{qq}(z,\eps) + I^{(3),\text{RRV,fin}}_{qq}(z,\eps) \,.
\end{align}
With our results for the master integrals we find that, for instance, the divergent part of the matching coefficient is given by the following compact expression
\begin{align}
\label{eq:Idiv}
&I^{(3),\text{RRV,div}}_{qq}(z,\eps) = 
\Re\left(e^{i \pi  \epsilon }\right) (1-z)^{-1 -3 \epsilon} \bigg\{
\nn\\&\hspace{0mm}
C_A C_F n_f 
\bigg[
\frac{1}{6 \epsilon }
+\frac{25}{18}
+\left(\frac{49}{9}+\frac{\pi ^2}{12}\right) \epsilon 
+\left(\frac{13 \zeta_3}{3}-\frac{\pi ^2}{4}+\frac{2777}{162}\right) \epsilon ^2
+ \mathcal{O}(\eps^3)
\bigg]
\nn\\&\hspace{-0.5mm}
+C_A C_F^2 
\bigg[
-\frac{9}{\epsilon ^4}
+\frac{9 \pi ^2}{2 \epsilon ^2}
+\frac{126 \zeta_3}{\epsilon }
+\frac{93 \pi ^4}{40}
+\left(810 \zeta_5-63 \pi ^2 \zeta_3\right) \epsilon 
+\left(\frac{53 \pi ^6}{80}-882 \zeta_3^2\right) \epsilon ^2
+ \mathcal{O}(\eps^3)
\bigg]
\nn\\&\hspace{-0.5mm}
+C_A^2 C_F 
\bigg[
-\frac{5}{2 \epsilon ^4}
-\frac{11}{2 \epsilon ^3}
+\frac{1}{\epsilon ^2}\left( \frac{19 \pi ^2}{12}-\frac{134}{9} \right)
+\frac{1}{\epsilon } \left( 63 \zeta _3+\frac{55 \pi ^2}{36}-\frac{1028}{27} \right)\nn\\&\hspace{17.5mm}
+\bigg(
\frac{1453 \pi ^4}{720}
+\frac{110 \zeta _3}{3}
+\frac{134 \pi ^2}{27}
-\frac{5093}{54}
\bigg)
+\bigg(
\frac{335 \zeta _3}{3}
-\frac{217 \pi ^2 \zeta _3}{6}
+775 \zeta _5 \nn\\&\hspace{17.5mm}
+\frac{187 \pi ^4}{720}
+\frac{2177 \pi ^2}{162}
-\frac{54418}{243}
\bigg) \epsilon 
+\bigg(
\frac{123427 \pi ^6}{90720}
-584 \zeta _3^2
-\frac{121 \zeta _5}{2}
-\frac{55 \pi ^2 \zeta _3}{18} \nn\\&\hspace{17.5mm}
+\frac{134 \pi ^4}{135}
+\frac{7648 \zeta _3}{27}
+\frac{35027 \pi ^2}{972}
-\frac{377464}{729}
\bigg) \epsilon ^2
+ \mathcal{O}(\eps^3)
\bigg]
\bigg\}
\,.
\end{align}
We remark that \cref{eq:Idiv} does not contain the colour factor $C_F^3$; this is similar to the observation
in ref.~\cite{Catani:2000pi} that eikonal factors are not renormalized by one-loop QED corrections.
To be clear, this does not imply the absence of $C_F^3$ in the full result for the $I_{qq}(t,z,\mu)$ matching coefficient, after all of its contributions have been included.


\section{Conclusions} 
\label{sec:concl}

We computed the master integrals for the double-gluon emission contribution to
the matching coefficient of the quark beam function in third-order perturbative QCD.
The matching coefficients are obtained as collinear limits of
QCD amplitudes, with additional constraints on the phase space that fix both light-cone
components of the real radiation.
We calculated the resulting non-standard phase-space integrals with the methods of
reverse unitarity and differential equations, and obtained the boundary conditions
from explicit computations of suitable integrals in the soft limit.
We provide  the master integrals as Laurent series in the
dimensional regulator $\epsilon$, which contain  multiple polylogarithms up to weight six.

The result of this paper is an important ingredient for the computation of the quark beam function through order $\mathcal{O}(\alpha_s^3)$ in QCD.
The completion of that task requires additionally the results for two-loop corrections to the single-real emission process as well as the triple-real emission process, which should be  feasible using  techniques employed  in this paper.

\appendix

\section{Loop integrals}
\label{sec:loopintegrals}

In this Appendix we collect all the one-loop integrals that are  required for computing the
virtual part of the double-real contribution to the matching coefficient of quark and gluon
beam functions. These one-loop integrals are given in a form that is convenient for evaluating
the required boundary constants as we explain in the main body of the paper. 

The one-loop bubble-type integrals are
\begin{align}
\text{Bub1}(q)
\equiv \int \frac{d^d k_3}{(2\pi)^{d}}~\frac{1}{k_{3}^2 \, (k_{3}-q)^2}
= C_{\text{bub}}(\eps) \left( - q^2 \right)^{-\eps} \,,
\label{eq:Bub1}
\end{align}
where the prefactor is defined as
$$C_{\text{bub}}(\eps) = \frac{i}{(4\pi)^{d/2}} \frac{\Gam(1-\eps)^2\Gam(\eps)}{\Gam(2-2\eps)}\,.$$

The one-loop two-mass triangle integral evaluates to
\begin{align}
\text{Tri1}
\equiv \int \frac{d^d k_3}{(2\pi)^d} \frac{1}{k_3^2 \, k_{13}^2 \, (p-k_{123})^2} 
= C_{\text{Tri}}(\eps) \, \frac{\left( - (p-k_2)^2 \right)^{-\eps} - \left( - (p-k_{12}) \right)^{-\eps}}{(p-k_2)^2 - (p-k_{12})^2}\,,
\label{eq:Tri1}
\end{align}
where the prefactor is defined as
$$C_{\text{Tri}}(\eps) = \frac{i}{(4\pi)^{d/2}}\, \frac{1}{\eps^2}\, \frac{\Gam(1-\eps)^2 \Gam(1+\eps)}{\Gam(1-2\eps)}\,.$$

The first one-loop one-mass box integral evaluates to \cite{Anastasiou:1999cx}
\begin{align}
\text{Box1} 
&\equiv \int \frac{d^d k_3}{(2\pi)^d} \frac{1}{k_3^2 \, k_{13}^2 \, k_{123}^2 \, (p-k_{123})^2} 
\nn\\
&= \frac{C_{\text{Box}}(\eps)}{k_{12}^2 \, (p-k_2)^2} \,
\bigg[\,
\left( -k_{12}^2 \right)^{-\eps} {}_2F_1\left(1,-\eps;1-\eps;\frac{-(p-k_1)^2}{(p-k_2)^2}\right) 
\nn\\
&\hspace{17mm} 
+\left( -(p-k_2)^2 \right)^{-\eps} {}_2F_1\left(1,-\eps;1-\eps;\frac{-(p-k_1)^2}{k_{12}^2}\right) 
\nn\\ 
&\hspace{-3mm} - \left( -(p-k_{12})^2 \right)^{-\eps} {}_2F_1\left(1,-\eps;1-\eps;\frac{-(p-k_1)^2\,(p-k_{12})^2}{k_{12}^2\,(p-k_2)^2}\right)
\bigg]\,,
\label{eq:Box1-v2}
\end{align}
where the prefactor is given by
$$C_{\text{Box}}(\eps) = \frac{i}{(4\pi)^{d/2}} \,\frac{2}{\eps^2}\, \frac{\Gam(1-\eps)^2 \Gam(1+\eps)}{\Gam(1-2\eps)}\,.
$$

The second one-loop one-mass box integral is
\begin{align}
\text{Box2} 
&\equiv \int \frac{d^d k_3}{(2\pi)^d} \frac{1}{k_3^2 \, k_{23}^2 \, (p-k_{23})^2 \, (p-k_{123})^2} 
\nn\\
&= \frac{C_{\text{Box}}(\eps)}{(p-k_2)^2 \, (p-k_1)^2} \,
\bigg[\,
\left( -(p-k_1)^2 \right)^{-\eps} {}_2F_1\left(1,-\eps;1-\eps;\frac{-k_{12}^2}{(p-k_2)^2}\right) \nn\\
&\hspace{38mm} +\left( -(p-k_2)^2 \right)^{-\eps} {}_2F_1\left(1,-\eps;1-\eps;\frac{-k_{12}^2}{(p-k_1)^2}\right) \nn\\
&\hspace{20mm} -\left( -(p-k_{12})^2 \right)^{-\eps} {}_2F_1\left(1,-\eps;1-\eps;\frac{-k_{12}^2\,(p-k_{12})^2}{(p-k_2)^2\,(p-k_1)^2}\right)
\bigg]\,.
\label{eq:Box2-v2}
\end{align}

The light-cone gauge propagator for the gluons contain the linear propagator $k_3.\bar{p}$. 
Because of partial fractioning, only one such denominator  appears in the master integrals. 
As a consequence, we encounter integrals with three, four and five propagators.
The triangle integral can be expressed in terms of the following hypergeometric function
\begin{align}
\text{Tri2}(q,m^2)
&\equiv \int \frac{d^d k_3}{(2\pi)^d} \frac{1}{(k_3)^2 (k_3+q)^2(k_3\cdot\bar{p}-m^2)} 
\nn\\
&= \widetilde{C}_{\text{Tri}}(\epsilon)\left(\frac{(-q^2)^{-\epsilon}}{m^2+q\cdot\bar{p}}\right) 
{}_2F_1\left(1-\epsilon,1,2-2\epsilon;\frac{q\cdot\bar{p}}{m^2+q\cdot\bar{p}}\right), 
\label{eq:Tri2-v2}
\end{align}
where the prefactor is given by $\widetilde{C}_{\text{Tri}}(\epsilon) = -\frac{i}{(4\pi)^{d/2}}\frac{\Gamma(1+\epsilon)\Gamma(1-\epsilon)^2}{\epsilon\Gamma(2-2\epsilon)}$.

The box integral is given as a  two-fold integral over Feynman parameters
\begin{align}
\text{Box3}(\tilde{p},q,m^2)
&\equiv \int \frac{d^d k_3}{(2\pi)^d} \frac{1}{(k_3)^2 (k_3+\tilde{p})^2 (k_3+q)^2(k_3\cdot\bar{p}-m^2)} 
\nn\\
&= 
\frac{i\Gamma(1+\epsilon)(-q^2)^{-1-\epsilon}}{(4\pi)^{d/2}}
\int_0^{\infty} dx_1 dx_2 
\nn\\&\hspace{10mm}
\frac{x_2^{-1-\epsilon}(1+x_1+x_2)^{2\epsilon}(1+(1-\frac{2q\cdot \tilde{p}}{q^2})x_1)^{-1-\epsilon}}{m^2(1+x_1+x_2)+x_1 \tilde{p}\cdot\bar{p}+x_2 q\cdot\bar{p}}\,,
\label{eq:Box3-v2}
\end{align}
where $\tilde{p}^2=0$.

Finally, the pentagon integral can be expressed through a three-fold integral over Feynman parameters
\begin{align}
\text{Pen1}(p_1,p_2,p_3)
&\equiv\int \frac{d^d k_3}{(2\pi)^d} \frac{1}{(k_3)^2 (k_3+p_1)^2 (k_3+p_{12})^2 (k_3+p_{123})^2(k_3\cdot\bar{p})} 
\nn\\
&= -\frac{i\Gamma(2+\epsilon)}{(4\pi)^{d/2}}\int_0^{\infty} dx_1 dx_2 dx_3 
\nn\\&\hspace{10mm}
\frac{(1+x_1+x_2+x_3)^{1+2\epsilon}(-s_{12} x_2-s_{123}x_3-s_{23}x_1x_3)^{-2-\epsilon}}{x_1 p_1\cdot\bar{p}+x_2 p_{12}\cdot\bar{p}+x_3p_{123}\cdot\bar{p}}\,,
\label{eq:Pen1-v2}
\end{align}
where $p_i^2=0,\, s_{12}=p_{12}^2,\, s_{23}=p_{23}^2,\, s_{123}=p_{123}^2$.

\acknowledgments
We thank A.~von Manteuffel for useful advice about the calculation of boundary integrals. 
We also thank C.~Duhr, P.~Monni, W.~Waalewijn and D.~Wellmann for useful discussions. 
LT wishes to thank C.~Duhr for giving him access to his package for the manipulation of multiple polylogarithms. 
The research of LT was supported by the ERC starting grant 637019 ``MathAm''. 
KM, LT and CW wish to thank the MIAPP in Munich, where part of this work was carried out.

\bibliographystyle{JHEP}
\bibliography{biblio}

\end{document}